\documentclass[iop]{emulateapj}

\usepackage{natbib,aas_macros,amsmath}
\usepackage{color}
\usepackage{longtable}
\usepackage{here}

\begin{document}
\slugcomment{Submitted to ApJ}
\shorttitle{Subaru/HSC Protoclusters at $z=6-7$}
\shortauthors{Higuchi et al.}

\title{
SILVERRUSH. VII. Subaru/HSC Identifications of 42 Protocluster Candidates at $\lowercase{z}\sim 6-7$ with the Spectroscopic Redshifts up to $\lowercase{z}=6.574$: \\Implications for Cosmic Reionization
}

\author{
Ryo Higuchi\altaffilmark{1,2}, 
Masami Ouchi\altaffilmark{1,3}, 
Yoshiaki Ono\altaffilmark{1}, 
Takatoshi Shibuya\altaffilmark{1}, 
Jun Toshikawa\altaffilmark{1}, 
Yuichi Harikane\altaffilmark{1,2},
Takashi Kojima\altaffilmark{1,2},
Yi-Kuan Chiang\altaffilmark{4},
Eiichi Egami\altaffilmark{5},
Nobunari Kashikawa\altaffilmark{6,7}, %
Roderik Overzier\altaffilmark{8}, 
Akira Konno\altaffilmark{1,9},\\ 
Akio K. Inoue\altaffilmark{10}, 
Kenji Hasegawa\altaffilmark{11}, 
Seiji Fujimoto\altaffilmark{1,9}, 
Tomotsugu Goto\altaffilmark{12}, 
Shogo Ishikawa\altaffilmark{13,14},\\ %
Kei Ito\altaffilmark{7},
Yutaka Komiyama\altaffilmark{6,7},
and
Masayuki Tanaka\altaffilmark{6}
}
\email{rhiguchi@icrr.u-tokyo.ac.jp}
\altaffiltext{1}{Institute for Cosmic Ray Research, The University of Tokyo, 5-1-5 Kashiwanoha, Kashiwa, Chiba 277-8582, Japan}
\altaffiltext{2}{Department of Physics, Graduate School of Science, The University of Tokyo, 7-3-1 Hongo, Bunkyo-ku, Tokyo 113-0033, Japan}
\altaffiltext{3}{Kavli Institute for the Physics and Mathematics of the Universe (Kavli IPMU, WPI), The University of Tokyo, 5-1-5 Kashiwanoha, Kashiwa, Chiba 277-8583, Japan}
\altaffiltext{4}{Department of Physics \& Astronomy, Johns Hopkins University, 3400 N. Charles Street, Baltimore, MD 21218, USA}
\altaffiltext{5}{Steward Observatory, University of Arizona, 933 North Cherry Avenue, Tucson, AZ 85721, USA}
\altaffiltext{6}{Optical and Infrared Astronomy Division, National Astronomical Observatory, Mitaka, Tokyo 181-8588, Japan.}
\altaffiltext{7}{Department of Astronomical Science, Graduate University for Advanced Studies (SOKENDAI), Mitaka, Tokyo 181-8588, Japan.}
\altaffiltext{8}{Observat$\acute{\rm o}$rio Nacional, Rua Jos$\acute{\rm e}$ Cristino, 77. CEP 20921-400, S$\tilde{\rm a}$o Crist$\acute{\rm o}$v$\tilde{\rm a}$o, Rio de Janeiro-RJ, Brazil}
\altaffiltext{9}{Department of Astronomy, Graduate School of Science, The University of Tokyo, 7-3-1 Hongo, Bunkyo-ku, Tokyo 113-0033, Japan}
\altaffiltext{10}{Department of Environmental Science and Technology, Faculty of Design Technology, Osaka Sangyo University, 3-1-1, Nakagaito, Daito 574-8530 Osaka, Japan}
\altaffiltext{11}{Department of Physics and Astrophysics, Nagoya University Furo-cho, Chikusa-ku, Nagoya, Aichi 464-8602, Japan}
\altaffiltext{12}{Institute of Astronomy, National Tsing Hua University, No. 101,Section 2, Kuang-Fu Road, Hsinchu, Taiwan}
\altaffiltext{13}{Center for Computational Astrophysics, National Astronomical Observatory of Japan, Mitaka, Tokyo 181-8588, Japan}
\altaffiltext{14}{Department of Science, Faculty of Science and Engineering, Kindai University, Higashi-Osaka, Osaka 577-8502, Japan}

\begin{abstract}
We report fourteen and twenty-eight protocluster candidates at $z=5.7$ and $6.6$ over 14 and 19 deg$^2$ areas, respectively, selected from 2,230 (259) Ly$\alpha$ emitters (LAEs) photometrically
(spectroscopically) identified with Subaru/Hyper Suprime-Cam (HSC) deep images 
(Keck, Subaru, and Magellan spectra and the literature data).
Six out of the 42 protocluster candidates include
$1-12$ spectroscopically confirmed LAEs at redshifts up to $z=6.574$. 
By the comparisons with the cosmological Ly$\alpha$ radiative transfer (RT) model reproducing LAEs 
with the reionization effects, we find that more than a half of these protocluster candidates are 
progenitors of the present-day clusters with a mass of $\gtrsim 10^{14} M_\odot$. 
We then investigate the correlation between LAE overdensity $\delta$ and Ly$\alpha$ rest-frame equivalent width $EW_{\rm Ly\alpha}^{\rm rest}$, because the cosmological Ly$\alpha$ RT model suggests that a slope of $EW_{\rm Ly\alpha}^{\rm rest}$-$\delta$ relation is steepened towards the epoch of cosmic reionization (EoR), due to the existence of the ionized bubbles around galaxy overdensities easing the escape of Ly$\alpha$ emission from the partly neutral intergalactic medium (IGM). 
The available HSC data suggest that the slope of the $EW_{\rm Ly\alpha}^{\rm rest}$-$\delta$ correlation 
does not evolve from the post-reionization epoch $z=5.7$ to the EoR $z=6.6$ beyond the moderately large statistical errors. 
There is a possibility that we would detect the evolution of the $EW^{\rm rest}_{\rm Ly\alpha}$ - $\delta$ relation 
from $z=5.7$ to $7.3$ by the upcoming HSC observations providing large samples of LAEs at $z=6.6-7.3$.
\end{abstract}


\section{Introduction} \label{sec:Introduction}
Studying the physical process of cosmic reionization is one of the important subjects in astronomy today. 
It is suggested that cosmic reionization was completed by $z\sim 6$ by the studies of 
the Gunn-Peterson effect and the Ly$\alpha$ damping wing
found in the continua of high-redshift QSOs and gamma-ray bursts \citep{fan2006,bolton2011,goto2011,chornock2013,mcgreer2015}. 
Similarly, Ly$\alpha$ emission in high-redshift galaxies is used to investigate the ionization state of the intergalactic 
medium (IGM), because the Ly$\alpha$ damping wing of HI gas in the IGM attenuates 
Ly$\alpha$ photons from Ly$\alpha$ emitters (LAEs). 
Recently, \cite{konno2017} and \cite{ouchi2017} have constrained the neutral hydrogen fraction of the IGM 
to be $x_{\rm HI}\simeq 0.3 \pm 0.2$ at $z=6.6$ from the evolution of the Ly$\alpha$ luminosity functions (LFs) 
and the angular correlation function based on the large samples of LAEs
at $z\sim 6-7$ (see also \citealt{malhotrarhoads2004,kashikawa2006,ouchi2008,ouchi2010,ota2010,konno2014}). 

Despite the fact that the mean values of $x_{\rm HI}$ at $z\sim 6-7$ are constrained, it is still unclear 
what are the ionizing photon sources of cosmic reionization. 
Although there are several candidates for the major sources of cosmic reionization, many observations suggest that it is likely that star-forming galaxies are major sources of cosmic reionization \citep{robertson2015,bouwens2015,ishigaki2017}. 
In this case, theoretical models predict that star-forming galaxies emitting ionizing photons from young massive stars 
would ionize the IGM around galaxies and that the ionized regions in the IGM are called ionized bubbles. 
Large ionized bubbles are expected to form in galaxy overdense regions, where many star-forming galaxies exist 
in a small volume of the universe \citep{furlanetto2006,ono2012,matthee2015,ishigaki2016,overzier2016,chiang2017}. 
The cosmic reionization is expected to proceed from high- to low-density regions 
(see \citealt{iliev2006,ono2012,overzier2016}). This reionization process is called 'inside-out scenario'. 
On the other hand, if major sources of cosmic reionization are X-ray emitting objects like AGNs, 
the scenario may be different.  Due to the longer mean-free path of X-ray photons than that of UV photons from galaxies
and the slow hydrogen recombination rate in the low-density region, cosmic reionization would 
not proceed from high-density, but low-density regions (see \citealt{miralda2000,nakamoto2001,mcquinn2012,mesinger2013}).  
The physical process of cosmic reionization is tightly related to the major ionizing sources of cosmic reionization. 
Because no definitive observational evidence of ionized bubbles is found to date,
identifying signatures of ionized bubbles around galaxy overdense regions, if any,
is key to testing the inside-out scenario of cosmic reionization. 

There is another importance of observations of galaxy overdensities near the EoR.
Standard structure formation models predict that a large fraction of high-$z$ galaxy overdense regions 
evolve into massive galaxy clusters at $z=0$. These galaxy overdense regions are called protoclusters. 
A protocluster is often defined as a structure expected to collapse into a galaxy cluster 
with a halo mass $M_{\rm h}>10^{14}\:\rm M_{\odot}$ \citep{chiang2013,overzier2016}. 
Galaxy overdensities at the EoR would be examples of the first site of the galaxy cluster formation 
(e.g. \citealt{ishigaki2016}).

Although the importances of high-$z$ galaxy overdensities are well recognized, 
only a few protoclusters at $z \gtrsim 6$ are reported, to date \citep{ouchi2005,utsumi2010,
toshikawa2012,toshikawa2014,franckmcgaugh2016a,franckmcgaugh2016b,
chanchaiworawit2017,toshikawa2017}. 
It is popular that protoclusters are identified with the distributions of the continuum-selected galaxies
including dropout galaxies.
However, there is a difficulty to find protoclusters only with the continuum-selected galaxy samples
due to the large redshift uncertainties of the continuum-selected galaxies.
Instead, one can use LAEs to identify protoclusters or galaxy overdensities in general,
exploiting a small redshift uncertainty of LAEs.
Here we investigate the LAE distribution and overdensity 
to identify protocluster candidates, and to
investigate the IGM ionization state around galaxy overdensities.
The IGM ionization state is studied with the Ly$\alpha$ equivalent widths (EWs) of LAEs
that depend on $x_{\rm HI}$ \citep{dijkstra2011,dijkstra2016,jensen2014,kakiichi2016}. 
Having a number of galaxy overdensities, we statistically investigate 
protoclusters and the IGM ionization states.

In this paper, we identify protocluster candidates at $z=5.7$ and $6.6$ based on the LAE samples of {\it Systematic Identification of LAEs for Visible Exploration and Reionization Research Using Subaru HSC (SILVERRUSH}; \citealt{ouchi2017}). 
SILVERRUSH is an on-going research project based 
on the Subaru/Hyper Suprime-Cam (HSC) Subaru Strategic Program (SSP; \citealt{aihara2017b}, \citealt{miyazaki2017}, \citealt{komiyama2017}, \citealt{furusawa2017}). 
The SHILVERRUSH project papers show various properties of LAEs in the EoR, 
clustering \citep{ouchi2017}, photometry \citep{shibuya2017a}, spectroscopy \citep{shibuya2017b}, Ly$\alpha$ LFs \citep{konno2017}, the ISM properties \citep{harikane2017b}, theoretical predictions \citep{inoue2017},
and protoclusters (this work). This is the seventh publication in SILVERRUSH. 
SILVERRUSH is one of the twin programs devoted to scientific results on high redshift galaxies 
based on the HSC survey data. The other one is related to dropout galaxies, named 
Great Optically Luminous Dropout Research Using Subaru HSC 
(GOLDRUSH; \citealt{ono2017}, \citealt{harikane2017a}, \citealt{toshikawa2017}). 
Because we intend to enlarge our LAE samples, we include the LAE samples made 
in \cite{ouchi2008} and \cite{ouchi2010}, which are previously obtained 
with Subaru/Suprime-Cam (SC; \citealt{miyazaki2002}; see also \citealt{iye2004}). %
We describe our photometric LAE samples with HSC and SC in Section \ref{sec:Data}. In Sections \ref{sec:spec} and \ref{sec:theo}, we explain our spectroscopic LAE data and theoretical models of \cite{inoue2017}, respectively. 
We present the list of protocluster candidates at $z=5.7$ and $6.6$, and show the 3-dimensional LAE distributions of protocluster candidates (Section \ref{sec:results}). 
In Section \ref{sec:results}, we also discuss the physical process of cosmic reionization with the LAE distributions.

Throughout this paper, we use a cosmological parameter set of $\Omega_\mathrm{m} = 0.3$, $\Omega_\mathrm{\Lambda} = 0.7$, $\Omega_\mathrm{b} = 0.04$, and $H_0 = 70$ km s$^{-1}$ Mpc$^{-1}$. 
The magnitudes are in the AB system.

\section{Data and Samples} \label{sec:Data}

\subsection{Photometric Samples of HSC SSP Data}\label{sec:hsclae}
We calculate galaxy overdensity and identify protocluster candidates using photometric LAE samples of HSC SSP data. 
In our study, we use two-narrowband ($NB816$ and $NB921$) and five-broadband ($grizy$) imaging data, 
of the HSC SSP survey (Section \ref{sec:Introduction}) starting in March 2014. 
The HSC-SSP survey is an on-going program, for which 300 nights are allocated over 5 years. 
The HSC-SSP survey has three layers of the UltraDeep, Deep, and Wide, whose 
planned total survey areas are $\sim$ 4 $\rm deg^2$, $\sim$ 30 $\rm deg^2$, and $\sim 1400$ $\rm deg^2$, respectively. 
The narrowband data are taken only in the UltraDeep and Deep layers. 
We use early datasets of the HSC-SSP survey taken until April 2016. 
In these datasets, HSC SSP has obtained $NB816$ data 
in two fields of the UltraDeep layer, UD-SXDS and UD-COSMOS, and two fields of the Deep layer, D-ELAIS-N1, 
and D-DEEP2-3. The data of $NB921$ have been taken in
two fields of the UltraDeep layer, UD-SXDS and UD-COSMOS, 
and three fields of the Deep layer, D-ELAIS-N1, D-DEEP2-3, and D-COSMOS. 
The $5\sigma$ limiting magnitudes of the HSC imaging data are typically $\simeq 25-25.5$ magnitudes 
in the narrowbands and $\simeq 26-27$ magnitudes in the broadbands (Table \ref{tab:photodata};
see also \citealt{shibuya2017a}). 
The total survey areas of the early datasets are 13.8 $\rm deg^2$ and 21.2 $\rm deg^2$ in the fields
with the $NB816$ and $NB921$ data, respectively. The $NB816$ and $NB921$ data
allow us to identify strong Ly$\alpha$ emission lines of LAEs redshifted to $z=5.726 \pm 0.046$ and $z=6.580 \pm 0.056$,
respectively, where the redshift ranges are defined with the FWHMs of the narrowbands.
The total survey volumes for the early datasets are $1.2 \times 10^7 {\rm \:Mpc}^3$ at $z=5.7$
and $1.9 \times 10^7 {\rm \:Mpc}^3$ at $z=6.6$.
Note that these survey volumes are $\sim 2-50$ and $\sim 4-100$ times larger than those of previous studies 
for LAEs at $z=5.7$ (e.g. \citealt{ouchi2008,santos2016}) and $z=6.6$ 
(e.g. \citealt{ouchi2010,kashikawa2011,matthee2015}), respectively. 

The datasets are reduced by the HSC-SSP Collaboration with $\tt{hscPipe}$ \citep{bosch2017}. 
$\tt{hscPipe}$ is a pipeline which is based on the Large Synoptic Survey Telescope (LSST) pipeline \citep{ivezic2008,axelrod2010,juric2015}. 
The astrometry and photometry of the datasets are calibrated based on the Panoramic Survey Telescope and Rapid Response System (Pan-STARRS) 1 imaging survey \citep{magnier2013,schlafly2012,tonry2012}. 

Our photometric samples of $z = 5.7$ and $6.6$ LAEs are made with combinations of the narrowband 
color excess and the UV continuum break \citep{shibuya2017a}. 
We apply color selection criteria which are similar to those of \cite{ouchi2008} and \cite{ouchi2010} 
who study $z=5.7$ and $6.6$ LAEs, respectively with SC. 
The color selection criteria to the objects in the HSC datasets are defined as

\begin{eqnarray}\label{eq:photo1}
\lefteqn{i - NB816 \geq  1.2 {\: \rm and \:} g > g_{3\sigma} {\: \rm and \:}} \nonumber \\
 &[(r { \leq } r_{3 \sigma} {\: \rm and\:} r - i {\geq} 1.0) {\: {\rm or} \:} (r > r_{3\sigma})]  
\end{eqnarray}
and \\
\begin{eqnarray}\label{eq:photo2}
\lefteqn{i - NB921 \geq 1.0 {\: \rm and \:} g > g_{3\sigma} {\: \rm and \:}r > r_{3\sigma}  {\: \rm and \:}  }\;\;\;\;\;\;\;\;\;\; \nonumber \\
 &[(z \leq z_{3\sigma} {\: \rm and \:} i - z \geq 1.0) {\: \rm or \:} (z > z_{3\sigma})]  
\end{eqnarray}
\\
for $z=5.7$ and $6.6$ LAEs, respectively (see \citealt{shibuya2017a}).
We find 1,077 $z=5.7$ LAEs and 1,153 $z=6.6$ LAEs by photometry.
\citet{shibuya2017b} take spectra of 18 LAEs of the photometric samples, and 
confirm 13 LAEs at $z=5.7$ and $6.6$ by spectroscopy. 
Because the LAEs include faint sources that may not identify a signal with the depth of the spectroscopy,
the contamination rate indicated by the spectroscopy is estimated to be 
$0-30$\% in the $z=5.7$ and $6.6$ LAEs in the photometric samples.

\begin{deluxetable*}{lccclcclll}[!t]
\tablecaption{HSC Imaging Data
\label{tab:photodata}}
\tablewidth{0pt}
\tablehead{
	\colhead{Layer}& \colhead{Field}& \colhead{Area}& \colhead{$g$}& \colhead{$r$}& \colhead{$i$}& \colhead{$z$}& \colhead{$y$}& \colhead{$NB816$}& \colhead{$NB921$}\\
	\colhead{(1)}& \colhead{(2)}& \colhead{$(3)$}& \colhead{$(4)$}& \colhead{$(5)$}& \colhead{$(6)$}& \colhead{$(7)$}& \colhead{$(8)$}& \colhead{$(9)$}& \colhead{(10)}
}
\startdata
UD & SXDS & 1.928 (1.873) & 26.9 & 26.4 & 26.3 & 25.6 & 24.9 & 25.5 & 25.5\\
UD & COSMOS & 1.965 (1.999) & 26.9 & 26.6 & 26.2 & 25.8 & 25.1 & 25.7 & 25.6\\ 
Deep & COSMOS & - (4.938) & 26.5 & 26.1 & 26.0 & 25.5 & 24.7 & - & 25.3\\
Deep & ELAIS-N1 & 5.566 (5.599) & 26.7 & 26.0 & 25.7 & 25.0 & 24.1 & 25.3 & 25.3\\
Deep & DEEP2-3 & 4.339 (3.100) & 26.6 & 26.2 & 25.9 & 25.2 & 24.5 & 25.2 & 24.9
\enddata
\tablecomments{
(1) Layer; 
(2) field; 
(3) effective area of the $NB816$ ($NB921$) image ($\rm deg^2$); 
(4)-(10) five sigma limiting magnitudes of the HSC $g$, $r$, $i$, $z$, $y$, $NB816$, and $NB921$ images
in a circular aperture with a diameter of $1.''5$ (mag).
}
\end{deluxetable*}
\begin{deluxetable*}{cccclcclll}
\tablecaption{Photometric Samples of the $z=5.7$ and $6.6$ LAEs
\label{tab:laesample}}
\tablewidth{0pt}
\tablehead{
	\colhead{}& \colhead{}& \colhead{$z=5.7$}& \colhead{}& \colhead{}&  \colhead{$z=6.6$}\\ \hline
	\colhead {Layer}& \colhead{Field} &  \colhead{$NB816$ Full}& \colhead{$NB816$ $<24.5^{\rm }$}& \colhead{$NB816$ $<25.0$}& \colhead{$NB921$ Full}& \colhead{$NB921$ $<24.5$}& \colhead{$NB921$ $<25.0$}&\\
	\colhead {(1)}& \colhead{(2)} &  \colhead{(3)}& \colhead{(4)}& \colhead{(5)}& \colhead{(6)}& \colhead{(7)}& \colhead{(8)}&	
}
\startdata
UD & SXDS & 224 & 83 & 164 & 58 & 21 & 43\\
UD & COSMOS & 201 & 52 & 123 & 338 & 31 & 82\\ 
Deep & COSMOS & -$^{\rm a}$ & -$^{\rm a}$ & -$^{\rm a}$ & 244 & 91 & 196\\
Deep & ELAIS-N1 & 229 & 140 & 166 & 349 & 142 & 258\\
Deep & DEEP2-3 & 423 & 127 & 319 & 164 & 104 & 82\\ \hline
 & Total & 1077 & 402 & 772 & 1153 & 389 & 661 
\enddata
\tablecomments{
(1) Layer;
(2) field;
(3) number of the $z=5.7$ LAEs in the HSC photometric sample; 
(4)-(5) same as (3), but for $z=5.7$ LAEs that are brighter than 24.5 and 25.0 mag in the $NB816$ band;
(6) number of the $z=6.6$ LAEs in the HSC photometric sample; 
(7)-(8) same as (6), but for $z=6.6$ LAEs that are brighter than 24.5 and 25.0 mag in the $NB921$ band. 
$^{\rm a}$ The $NB816$ image is not taken in Deep COSMOS. 
}
\end{deluxetable*}


\subsection{Photometric Samples of the SC Data}\label{sec:sclae}
To select the spectroscopic targets of $z=5.7$ and $6.6$ LAEs, we use photometric samples of \cite{ouchi2008} and \cite{ouchi2010}, respectively, in addition to the HSC LAE samples in Section \ref{sec:hsclae}. 
\cite{ouchi2008} and \cite{ouchi2010} have carried out narrowband imaging with SC in 2003 and 2005-2007, respectively. The total areas of the narrowband imaging are $\sim 1 \: \rm deg^2$ and $\sim 0.9 \: \rm deg^2$ for $NB816$ and $NB921$ images, respectively. \citet{ouchi2008} and \citet{ouchi2010}
detect objects in each narrowband image with SExtractor \citep{bertinarnouts1996},
and obtain SC LAE samples with the color selection criteria similar to the equations (\ref{eq:photo1}) and (\ref{eq:photo2})
that are defined as

\begin{eqnarray}
\lefteqn{i^{'} - NB816 \geq  1.2 {\: \rm and \:} B > B_{2\sigma} {\: \rm and \:}  V > V_{2\sigma} {\: \rm and \:} } \;\;\;\;\;\;\;\;\;\;\;\;\;\;\; \\
 &[(R {\leq} R_{2 \sigma} {\: \rm and\:} R - i^{'} {\geq} 1.0) {\: {\rm or} \:} (R > R_{2\sigma})] \nonumber
\end{eqnarray}

and \\

\begin{eqnarray}
\lefteqn{z^{'} - NB921 \geq 1.0 {\: \rm and \:} B > B_{2\sigma} {\: \rm and \:}V > V_{2\sigma}  {\: \rm and \:}  }\;\;\;\;\;\;\;\;\;\;\;\;\;\;\; \\
 &[(z^{'} \leq z^{'}_{3\sigma} {\: \rm and \:} i^{'}-z^{'} \geq 1.3) {\: \rm or \:} (z^{'} > z^{'}_{3\sigma})] \nonumber
\end{eqnarray}
\\
for $z=5.7$ and $6.6$ LAEs, respectively.  \citet{ouchi2008} and \citet{ouchi2010} apply 
these selection criteria, and find 401 and 207 LAEs at $z=5.7$ and $6.6$, respectively. 

Comparing the SC samples with the HSC samples, one can recognize that 
many LAEs in the SC samples are not included in the HSC samples.
This is because the SC samples have faint LAEs down to the narrowband magnitudes of $\sim 26$ mag,
while the depth of the HSC samples only reaches $\sim 25$ mag.


\section{Spectroscopic Observations and Samples} \label{sec:spec}
We conduct spectroscopic observations for the HSC and SC LAE samples.
The spectroscopic observations for the HSC samples are presented in \citet{shibuya2017b}.
Here we explain our spectroscopy for the SC samples that were conducted in 2007-2010.
\begin{figure}
	\plotone{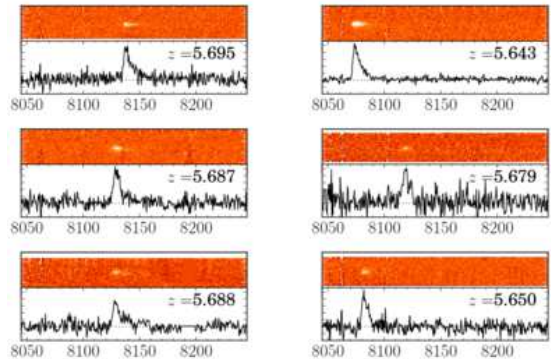}
	\caption{
Examples of the $z=5.7$ LAE spectra obtained by our Keck/DEIMOS observations. 
The two and one dimensional spectra are shown in the top and bottom subpanels, respectively,
in each panel.
The x-axis represents the observed wavelength. 
	}
	\label{fig:spdeimos}
\end{figure}
\begin{figure}
	\plotone{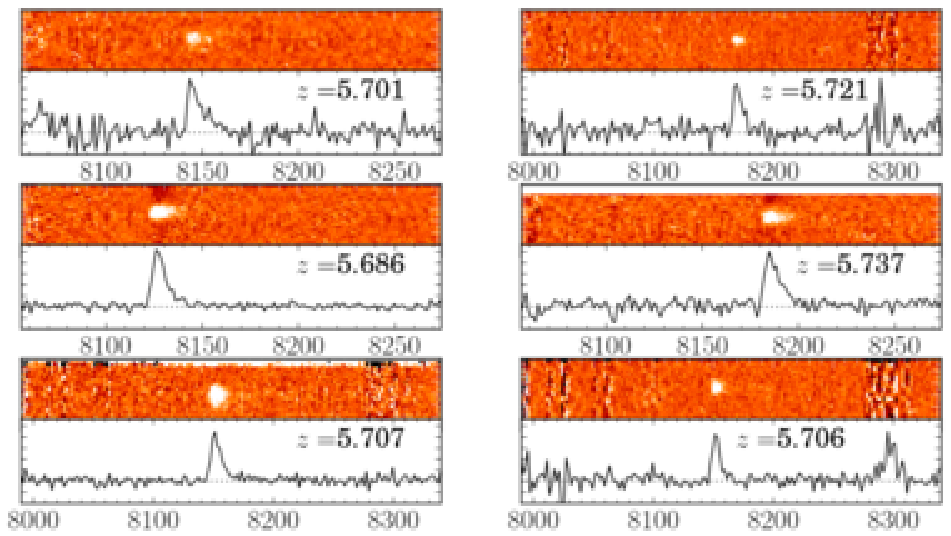}
	\caption{
Same as Figure \ref{fig:spdeimos}, but for our Magellan/IMACS spectra of the $z=5.7$ LAEs. 
	}
	\label{fig:spimacs}
\end{figure}
\begin{figure}
	\plotone{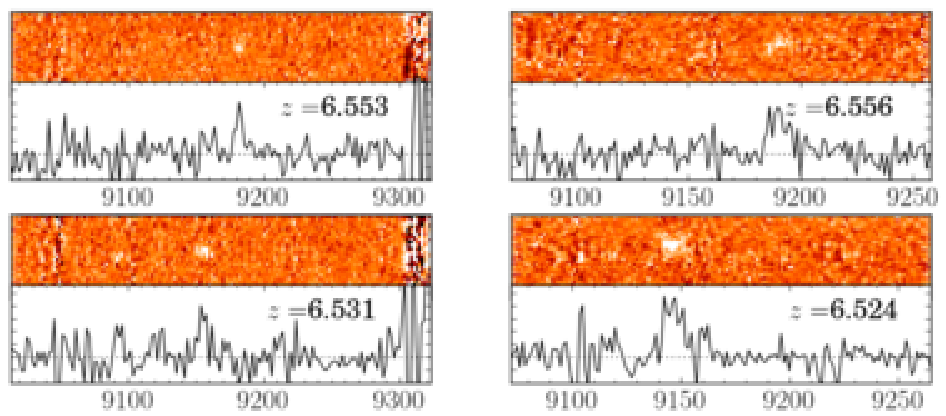}
	\caption{
Same as Figure \ref{fig:spdeimos}, but for our Magellan/IMACS spectra of the $z=6.6$ LAEs. 
	}
	\label{fig:spimacsn9}
\end{figure}


\begin{deluxetable*}{lllllllll}
\setlength{\tabcolsep}{6pt}
\tablecaption{Spectroscopic Observations
}
\label{tab:obsdata}
\tablewidth{0pt}
\tablehead{
	\colhead{Layer}& \colhead{Field}& \colhead{Mask ID}& \colhead{Date}& \colhead{Total Exposure}& \colhead{${N_{\rm LAE}}$}& \colhead{Grism}& \colhead{$\rm CW$}& \colhead{Filter} \\
	\colhead{(1)}& \colhead{(2)}& \colhead{(3)}& \colhead{(4)}& \colhead{(5)}& \colhead{(6)}& \colhead{(7)}& \colhead{(8)}& \colhead{(9)}
}
\startdata
	\multicolumn{7}{c}{Keck/DEIMOS}\\ \hline
UD & SXDS & SXDS03 & 2010 Feb 11 & 3500 & 16 & 830 & 7900 & OG550  \\ \hline
	\multicolumn{7}{c}{Magellan/IMACS}\\ \hline
UD & SXDS & sxds1\_07 & 2007 Nov 12 & 15600 & 4 & Gri-150-18.8 & - & GG455\\
UD & SXDS & sxds3r07 & 2007 Nov 13-14 & 37800 & 15 & Gri-150-18.8 & - & OG570\\

UD & SXDS & sxds5s08 & 2008 Nov 29 & 15300 & 22 & Gri-300-26.7 & - & WB6300-9500\\
UD & SXDS & sxds3s08 & 2008 Nov 30 & 15300 & 27 & Gri-300-26.7 & - & WB6300-9500\\
UD & SXDS & sxds2a08 & 2008 Dec 1 & 16200 & 22 & Gri-300-26.7 & - & WB6300-9500\\
UD & SXDS & sxds4a08 & 2008 Dec 2 & 18000 & 12 & Gri-300-26.7 & - & WB6300-9500\\

UD & SXDS & sxds1u08 & 2008 Dec 18-19 & 25200 & 11 & Gri-300-26.7 & - & WB6300-9500\\
UD & SXDS & sxds3a09 & 2009 Oct 11 & 21600 & 26 & Gri-300-26.7 & - & WB6300-9500\\
UD & SXDS & sxds2a09 & 2009 Oct 12 & 15300 & 16 & Gri-300-26.7 & - & WB6300-9500\\

UD & SXDS & $\rm oct2008\_nb1190a^a$ & 2008 Oct 20-23 & 22500 & 9 & Gri-300-26.7 & - & WB6300-9500\\
UD & SXDS & $\rm oct2008\_nb1190b^a$ & 2008 Oct 20-23 & 18000 & 4 & Gri-300-26.7 & - & WB6300-9500\\
UD & SXDS & $\rm oct2008\_nb1190c^a$ & 2008 Oct 20-23 & 18000 & 8 & Gri-300-26.7 & - & WB6300-9500\\
UD & SXDS & $\rm oct2008\_nb1190d^a$ & 2008 Oct 20-23 & 18000 & 11 & Gri-300-26.7 & - & WB6300-9500\\
UD & SXDS & $\rm sep2009\_sxdsw1^a$ & 2009 Sep 19-20 & 12600 & 1 & Gri-300-26.7 & - & WB6300-9500\\
UD & SXDS & $\rm sep2009\_sxdsw2^a$ & 2009 Sep 19-20 & 12600 & 3 & Gri-300-26.7 & - & WB6300-9500\\

UD & COSMOS & cos01\_08 & 2008 Nov 29 - Dec 2 & 21600 & 7 & Gri-300-26.7 & - & WB6300-9500\\
UD & COSMOS & cos02\_08 & 2008 Dec 18-20 & 27900 & 5 & Gri-300-26.7 & - & WB6300-9500
\enddata
\tablecomments{
(1) Layer; 
(2) field; 
(3) mask ID; 
(4) date of observations;
(5) total exposure time (sec);
(6) numbers of the observed LAEs; 
(7) disperser name; 
(8) central wavelength of the grating setting $(\rm \AA)$;
(9) filter name.
$\rm ^a$ See also \cite{lee2012} and \cite{momcheva2013}. 
}
\end{deluxetable*}

\subsection{Keck/DEIMOS Observation} \label{sec:spec_deimos}
We carried out spectroscopic follow-up observations for our $z=5.7$ LAEs with Deep Imaging Multi-Object Spectrograph (DEIMOS; \citealt{faber2003}) on 2010 February 11. 
The sky was clear during the observations, and the seeing was $\sim 0''.5$. 
We observed 22 out of the 401 SC LAEs at $z=5.7$ \citep{ouchi2008} including
very faint LAE candidates, 
and obtained 16 spectra in a good condition.
During the observations, we took the standard stars G191B2B for the flux calibration. 
We used a mask with a slit width of $1''$, the OG550 filter, and the 830 lines ${\rm mm}^{-1}$ grating 
that is blazed at 8640 \AA. 
The grating was tilted to be placed at a central wavelength of 7900 \AA \ on the detectors. 
The spectral coverage and the spectral resolution were $4100 - 9400$\AA\ 
and $\lambda/{\Delta}{\lambda} \simeq 2400$, respectively.

We perform the data reduction using the spec2d IDL pipeline 
developed by the DEEP2 Redshift Survey Team \citep{davis2003}. 
The central wavelengths of Ly$\alpha$ emission were determined by Gaussian fitting. 
We detect 15 out of the 16 LAEs, and
obtain Ly$\alpha$ line redshifts.
The spectra of the example LAEs are
shown in Figure \ref{fig:spdeimos}. 

\subsection{Magellan/IMACS Observation} \label{sec:spec_imacs}
We conducted follow-up spectroscopy for 425 objects selected from the samples of $z=5.7$ and $z=6.6$ LAEs in \cite{ouchi2008} and \cite{ouchi2010}, respectively. 
The observations were performed with the Inamori Magellan Areal Camera and Spectrograph (IMACS; \citealt{dressler2006}) on the Magellan I Baade Telescope in 2007 November $12-14$, 2008 November 29 $-$ December 2, 2008 December $18-19$, and 2009 October $11-12$. 
We chose GG455 filter and Gri-150-18.8 grism on 2007 November 12. 
In 2007 November $13-14$, we change the filter from GG455 to OG570. 
For the rest of the IMACS observations, we used WB6300-9500 filter and gri-300-26.7 grism. 
The exposure time ranges from 15,300 s to 35,400 s with seeing sizes of $0''.5 - 0''.8$. 
We used a $0''.8$ slit width that gives a spectral resolution of $1,000-2,000$. 
We perform data reduction with the Carnegie Observatories System for MultiObject Spectroscopy (COSMOS) pipeline, 
and detect Ly$\alpha$ emission lines around 8160 \AA\ (9210 \AA\ ) for 130 (22) objects.
Spectra of the example LAEs are shown in Figures \ref{fig:spimacs}-\ref{fig:spimacsn9}. 


\subsection{Spectroscopic Samples and Catalogs}
Adding to the SC spectroscopic sample of the LAEs 
confirmed with DEIMOS and IMACS 
in Sections \ref{sec:spec_deimos} and \ref{sec:spec_imacs}
and the HSC spectroscopic sample of \citealt{shibuya2017b} that includes LAEs in
\citealt{ouchi2010}, \citealt{sobral2015}, and \citealt{hu2016}, 
we use the redshift catalogs for the spectroscopically confirmed LAEs 
at $z=5.7$ ($6.6$) taken from \cite{ouchi2005}, \cite{ouchi2008}, \cite{mallery2012}, \cite{chanchaiworawit2017}, and \cite{guzman2017}.
We make unified spectroscopic catalogs of LAEs at $z=5.7$ and $6.6$
Tables \ref{tab:spec_n8uds} and \ref{tab:spec_n9uds}, respectively.

Note that, again, there are many LAEs in the SC spectroscopic sample
that are not included in the HSC photometric sample.
This is because the HSC photometric sample includes bright LAEs only
down to $\sim 25$ mag in a narrowband, while the SC samples 
(spectroscopic and photometric samples) have faint LAEs
down to $\sim 26$ mag in a narrowband (Section \ref{sec:sclae}).
Because the selection of the SC (and HSC) spectroscopic sample
is heterogeneous, we use the homogeneous photometric sample of HSC LAEs
to find protocluster candidates. The unified catalogs (the SC and HSC spectroscopic samples) 
are referred to confirm the redshifts of protocluster candidates in Section \ref{sec:three}. 
\begin{figure}
	\plotone{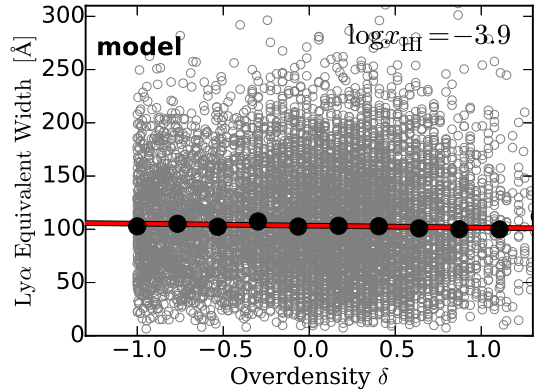}
	\caption{
$EW^{\rm rest}_{\rm Ly\alpha}$ as a function of $\delta$ at $\log_{\rm 10}{x_{\rm HI}}= -3.9$ (corresponding to $z=5.7$) 
in the model of \cite{inoue2017}. 
The values of $\delta$ and $EW^{\rm rest}_{\rm Ly\alpha}$ are shown with the gray open circles. 
The black filled circles and the bars indicate the median values and the error bars, respectively. 
The red line represents the best-fit linear function.
	}
	\label{fig:ewdelsimn8}
\end{figure}

\begin{figure}
	\plotone{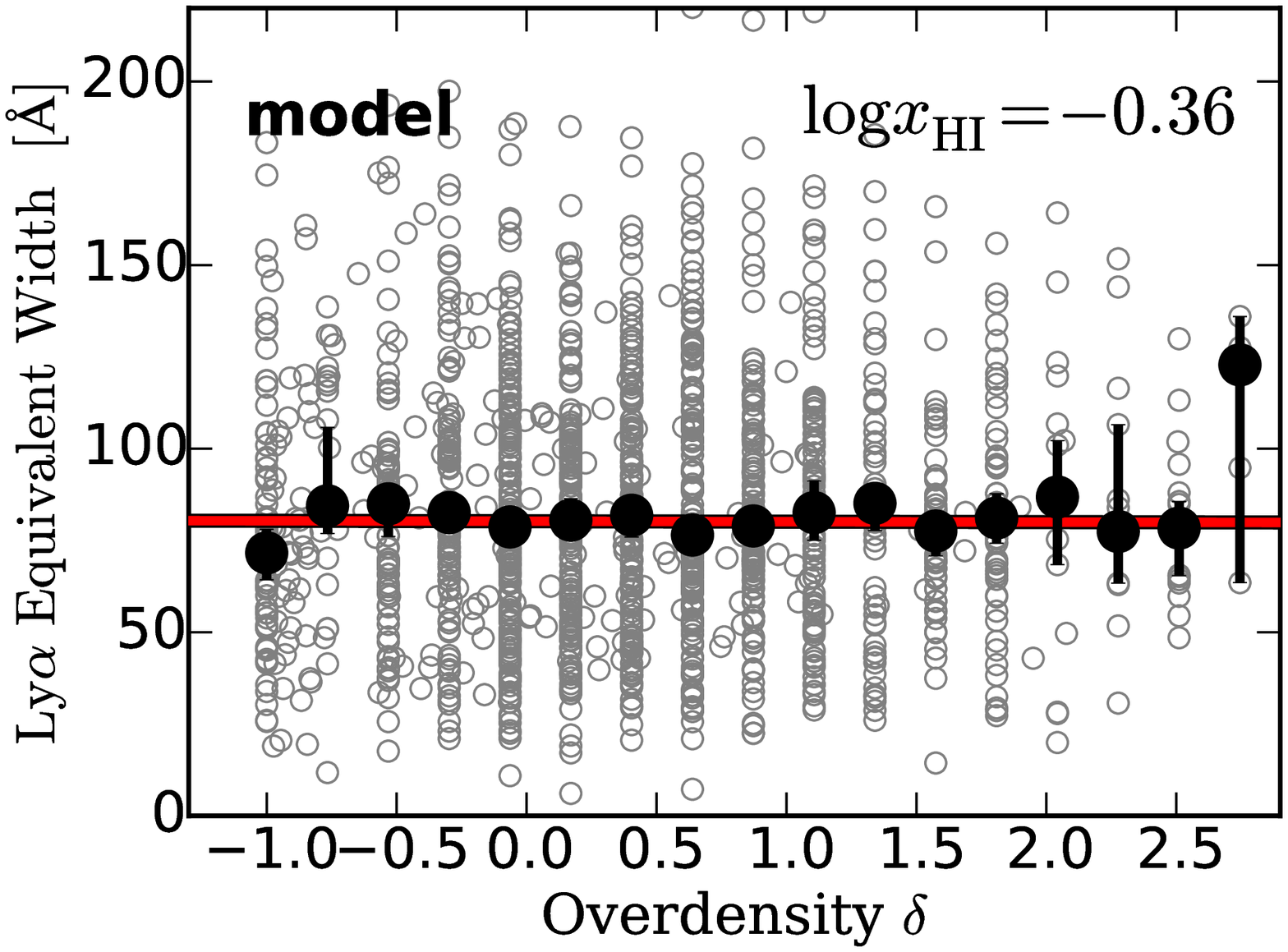}
	\caption{
Same as Figure \ref{fig:ewdelsimn8}, but for $\log_{\rm 10}{x_{\rm HI}}= -0.36$ (corresponding to $z=6.6$). 
	}
	\label{fig:ewdelsimn9}
\end{figure}

\begin{figure}
	\plotone{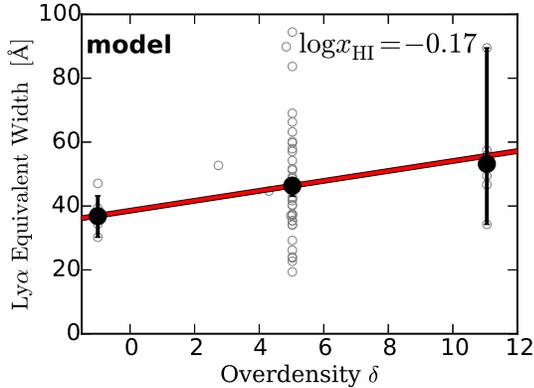}
	\caption{
Same as Figure \ref{fig:ewdelsimn8}, but for $\log_{\rm 10}{x_{\rm HI}}= -0.17$ (corresponding to $z=7.3$). 
	}
	\label{fig:ewdelsimn1}
\end{figure}

\begin{figure}
	\plotone{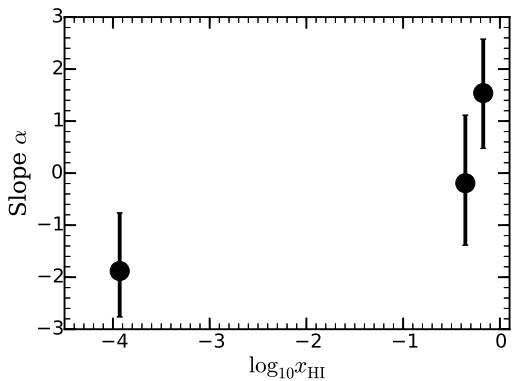}
	\caption{
Slope $\alpha$ of the $EW^{\rm rest}_{\rm Ly\alpha}$ - $\delta$ relation 
as a function of $x_{\rm HI}$ that is derived with the model of \cite{inoue2017}.
	}
	\label{fig: xhi_slope}
\end{figure}

\section{Theoretical Model} \label{sec:theo}

We compare our observational results with the cosmological simulation model of \citet{inoue2017}. 
\citet{inoue2017} conduct the N-body simulations in a box size of $110h^{-1}$ comoving Mpc (cMpc) length 
with $512^3$ grids, which gives a spatial resolution of $214.8$ comoving kpc. 
\citet{inoue2017} present models of three reionization histories depending on the ionizing emissivity of halos: 
$early$, $mid$, and $late$, all of which are consistent with the latest Thomson scattering optical depth measurement 
\citep{planck2016b}. Here we adopt the $late$ model that explains the recent neutral hydrogen fraction measurements
at $z\sim 6-7$. %
In the model, a total of $4096^3$ dark matter particles are used with a mass resolution of $7 \times 10^7$ $M_{\odot}$. 
\citet{inoue2017} perform numerical radiative transfer calculations to reproduce cosmic reionization.
In this model, LAEs are created with the relation of the Ly$\alpha$ photon production rate 
and halo mass determined by the radiation hydrodynamics (RHD) galaxy formation simulation 
of \citet{hasegawa2017}. \citet{inoue2017} assume 

\begin{equation}
L_{\rm Ly\alpha,int}=10^{42}\times(1-e^{-10 M_{\rm h,10}})\times{M_{\rm h,10}}^{1.1}\times 10^{\delta_{\rm Ly\alpha}} \: [\rm erg\:s^{-1}],
\end{equation}
where more massive haloes produce more Ly$\alpha$ photons due to the higher star-forming rate (SFR). 
Here, ${M_{\rm h,10}}$ is the halo mass normalized by $10^{10}M_{\odot}$, and 
$\delta_{\rm Ly\alpha}$ represents the fluctuation of the Ly$\alpha$ photon production. 
The ISM Ly$\alpha$ escape fraction is defined as 
\begin{eqnarray}
f_{{\rm esc},\alpha}^{\rm ISM}=\exp(-\tau_{\rm \alpha}),
\end{eqnarray}
where $\tau_{\alpha}$ is the Ly$\alpha$ optical depth. 
\citet{inoue2017} assume the probability distribution of the Ly$\alpha$ optical depth as

\begin{eqnarray}
P(\tau_{\alpha})=\frac{\exp\{-(\tau_{\alpha}-\langle \tau_{\alpha} \rangle)^2/2{\langle \tau_{\rm \alpha} \rangle}\}}{\sqrt{2\pi \langle \tau_{\alpha} \rangle}}
\end{eqnarray}

and

\begin{eqnarray}
\label{eq: tau}
\langle\tau_{\alpha}\rangle=\tau_{\alpha,10}\left(\frac{M_{\rm h}}{10^{10}M_{\odot}}\right)^p, 
\end{eqnarray}
where $p$ indicates the halo mass dependence of $\langle \tau_{\alpha}\rangle$. 
\cite{inoue2017} calibrate 
the parameter $\tau_{\alpha,10}$ with the $z=5.7$ Ly$\alpha$ luminosity function \citep{konno2017}, 
and compare the model predictions with the various observational quantities of
the Ly$\alpha$ luminosity functions at $z=6.6$ and $7.3$ \citep{konno2017,konno2014}, 
the LAE angular auto-correlation functions at $z=5.7$ and $6.6$ \citep{ouchi2017}, 
and the LAE fractions in Lyman break galaxies at $z=5-7$ \citep{stark2011,ono2012}. 
In this paper, we use the model with the best parameter set 
($\delta_{\rm Ly\alpha}=0$, $p=1/3$, and $\tau_{\alpha,10}=1.1$)
that \cite{inoue2017} conclude.

We select mock LAEs brighter than $10^{42.5}\:\rm erg\:s^{-1}$ in Ly$\alpha$ luminosity. 
Hereafter, we call these mock LAEs 'LAE all'. 
We obtain 9574, 1415, and 55 mock LAEs at $z=5.7$, $6.6$, and $7.3$, respectively,
from the entire simulation box of the model.

For comparison with our observational results,
we calculate overdensity $\delta$ of the mock LAEs 
that is defined as 

\begin{eqnarray}
\label{eq:delta}
\delta=\frac{n-\overline{n}}{\overline{n}},
\end{eqnarray}
where $n$ ($\overline{n}$) is the total (average) number of LAEs found in a cylinder volume
that mimicks the observational volume for the $\delta$ measurements (Section \ref{sec:ovme}).
We choose the height of $\sim40$ cMpc for the cylinder 
that corresponds to the redshift range of the narrowband observation LAE selection. 
The base area of the cylinder is defined by a radius of 10 cMpc that is a typical size of
protoclusters assumed in \citet{chiang2013,lovell2017}. 
Figures \ref{fig:ewdelsimn8}-\ref{fig:ewdelsimn1} show the relations between Ly$\alpha$ rest-frame equivalent width $EW^{\rm rest}_{\rm Ly\alpha}$ and $\delta$ in \cite{inoue2017} for the universe 
with the neutral hydrogen fractions of $\log_{10}{x_{\rm HI}}$ $=-3.9$, $-0.36$ and $-0.17$
that are the average values of the simulation boxes at $z=5.7$, $6.6$, and $7.3$, respectively.
The relations of $EW^{\rm rest}_{\rm Ly\alpha}-\delta$ are fit 
with a linear function, $EW^{\rm rest}_{\rm Ly\alpha}=\alpha \delta+EW_{\delta=0}$, where $\alpha$ and $EW_{\delta=0}$ 
are the slope and the $EW^{\rm rest}_{\rm Ly\alpha}$ value at $\delta=0$, respectively. 
Figure \ref{fig: xhi_slope} shows $\alpha$ as a function of $x_{\rm HI}$ obtained by the model calculations.
The slope $\alpha$ increases from the post reionization epoch ($\log_{10}{x_{\rm HI}} = -3.9$)
to the EoR ($\log_{10}{x_{\rm HI}}=-0.36$ and $-0.17$).
In the inside-out scenario of cosmic reionization, 
$EW^{\rm rest}_{\rm Ly\alpha}$ values at high-overdensity regions 
would be higher than those at lower-overdensity regions. 
This is because the Ly$\alpha$ escape fraction is higher inside the ionized bubbles than outside the ionized bubbles. 
Thus, if cosmic reionization proceeds in the inside-out manner, a slope $\alpha$ is high at the EoR. 


\begin{figure}
	\plotone{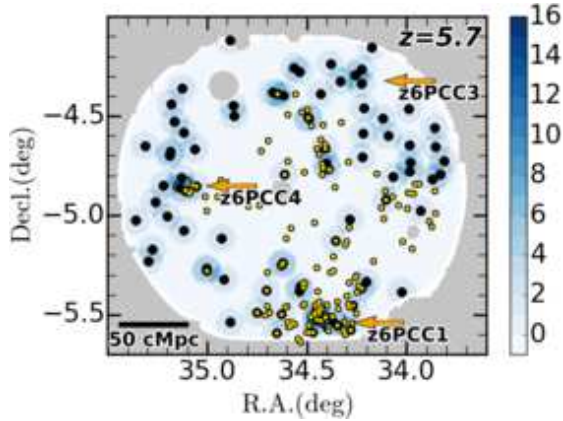}
	\caption{
Sky distribution of the $z=5.7$ LAEs (black filled circles) with the surface overdensity contours (black lines) 
in the $NB816$ UD-SXDS field. 
Higher density regions are shown with bluer colors.
The yellow filled circles represent spectroscopically-confirmed LAEs. 
Extended Ly$\alpha$ sources found in \cite{shibuya2017a} are marked with the black filled squares. 
The black lines correspond to the 5$\sigma$ to 8$\sigma$ significance levels of overdensity $\delta$ 
with a step of 1$\sigma$. Masked regions are presented with gray regions. 
	}
	\label{fig:coln8uds}
\end{figure}

\begin{figure}
	\plotone{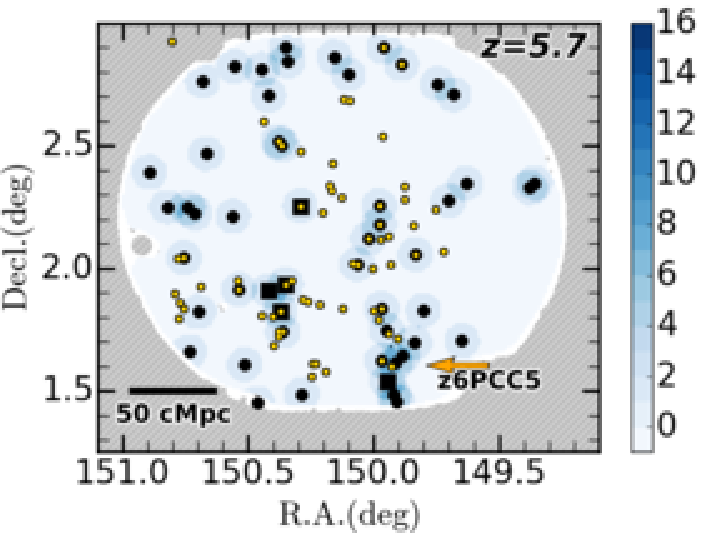}
	\caption{
Same as Figure \ref {fig:coln8uds}, but for the $z=5.7$ LAEs in UD-COSMOS field.
	}
	\label{fig:coln8udc}
\end{figure}
\begin{figure*}
	\plotone{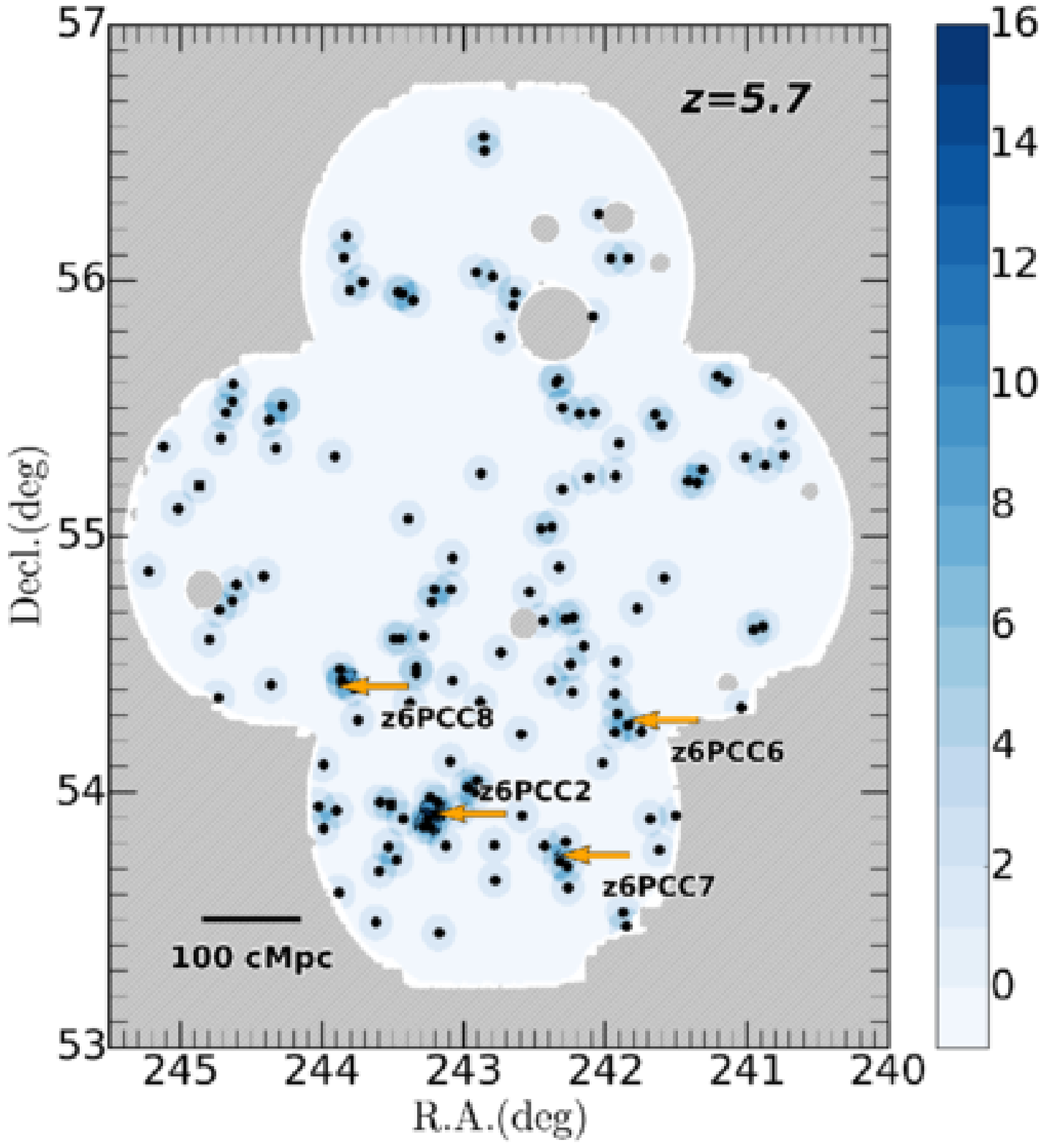}
	\caption{
Same as Figure \ref{fig:coln8uds}, but for the $z=5.7$ LAEs in D-ELAIS-N1 field.
	}
	\label{fig:coln8dde}
\end{figure*}

\section{Results and Discussion}\label{sec:results}

\begin{figure*}
	\plotone{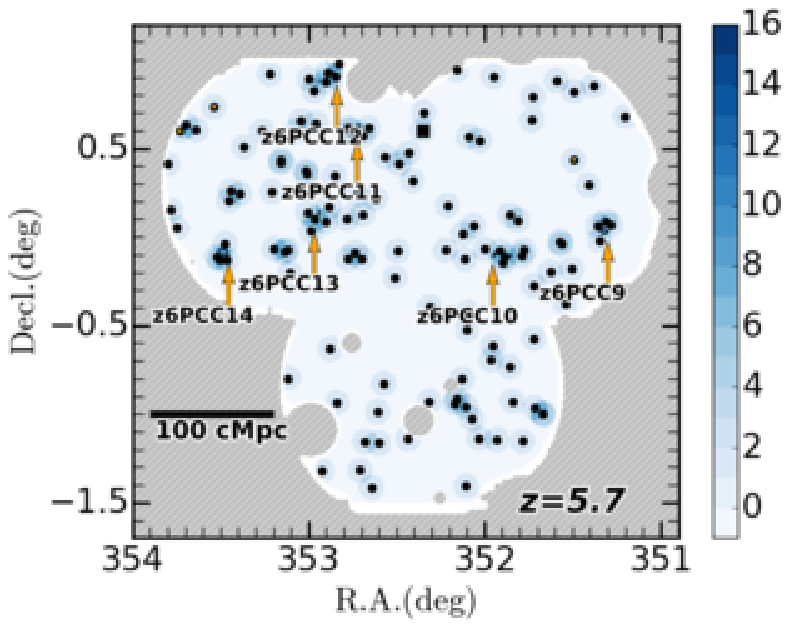}
	\caption{
Same as Figure \ref{fig:coln8uds}, but for the $z=5.7$ LAEs in D-DEEP2-3 field.
	}
	\label{fig:coln8del}
\end{figure*}


\subsection{Spatial Distribution of LAEs}\label{sec:spa}
\subsubsection{Overdensity Measurements}\label{sec:ovme}
\begin{figure}
	\plotone{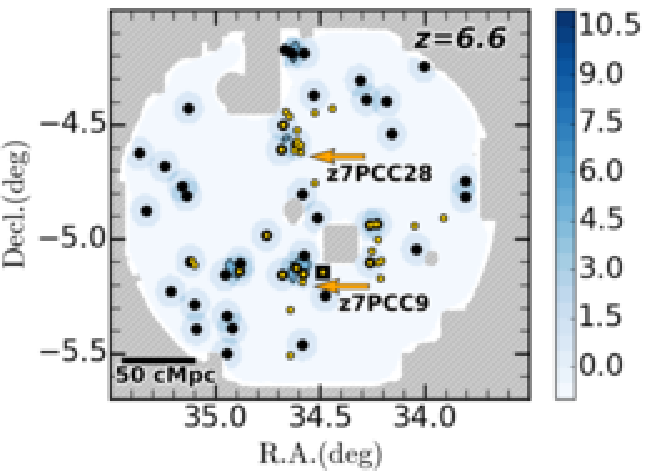}
	\caption{
Same as Figure \ref{fig:coln8uds}, but for the $z=6.6$ LAEs in UD-SXDS field. 
The solid lines correspond to contours of $\delta$ from $3 \sigma$ to $7 \sigma$ significance levels with a step of 1$\sigma$. 
	}
	\label{fig:coln9uds}
\end{figure}

\begin{figure}
	\plotone{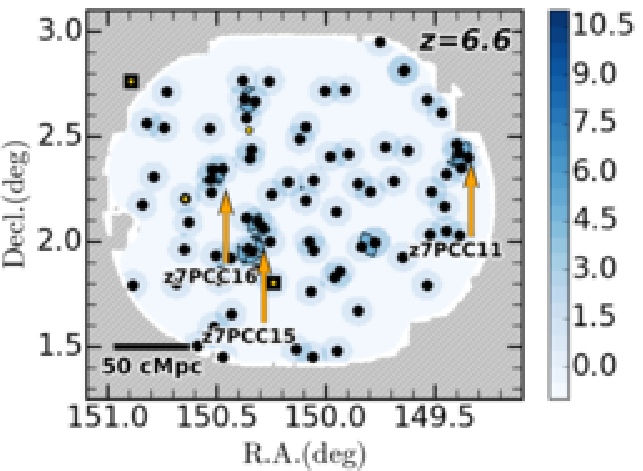}
	\caption{
Same as Figure \ref{fig:coln8uds}, but for the $z=6.6$ LAEs in UD-COSMOS field.
	}
	\label{fig:coln9udc}
\end{figure}

\begin{figure*}
	\plotone{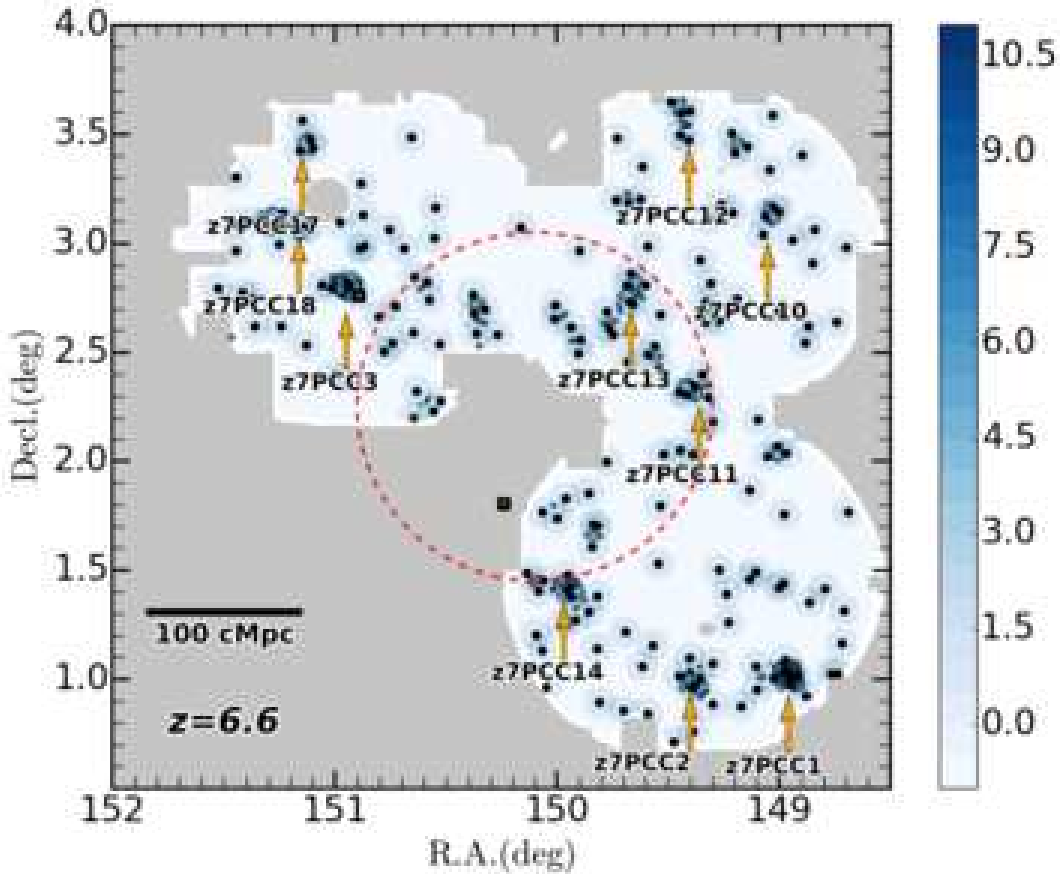}
	\caption{
Same as Figure \ref{fig:coln8uds}, but for the $z=6.6$ LAEs in D-COSMOS field. 
The red-dashed line represents the region of the $z=6.6$ LAE UD-COSMOS field that is shown 
in Figure \ref{fig:coln9udc}. 
	}
	\label{fig:coln9del}
\end{figure*}

\begin{figure*}
	\plotone{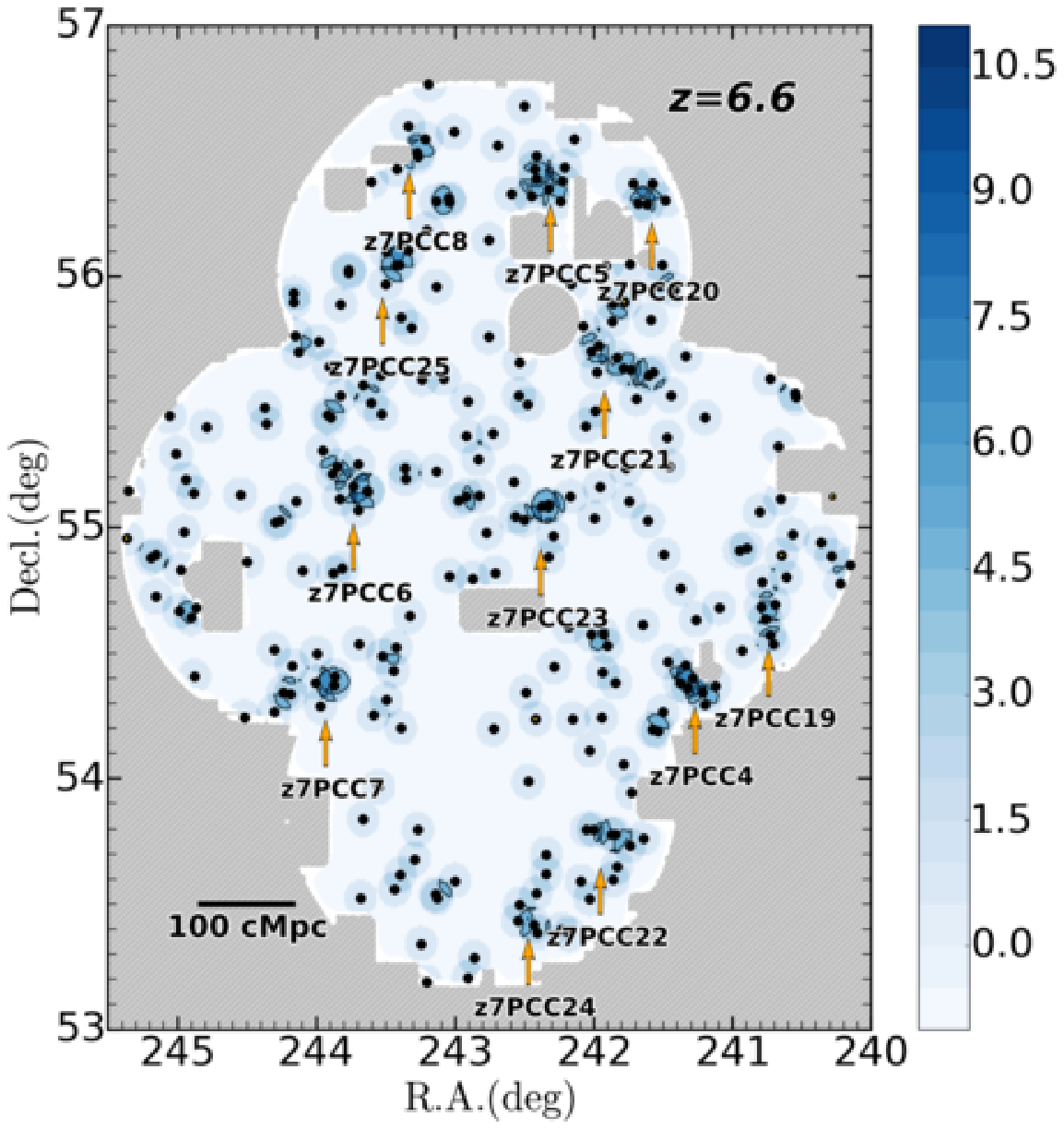}
	\caption{
Same as Figure \ref{fig:coln8uds}, but for the $z=6.6$ LAEs in D-ELAIS-N1 field.
	}
	\label{fig:coln9dde}
\end{figure*}

\begin{figure*}
	\plotone{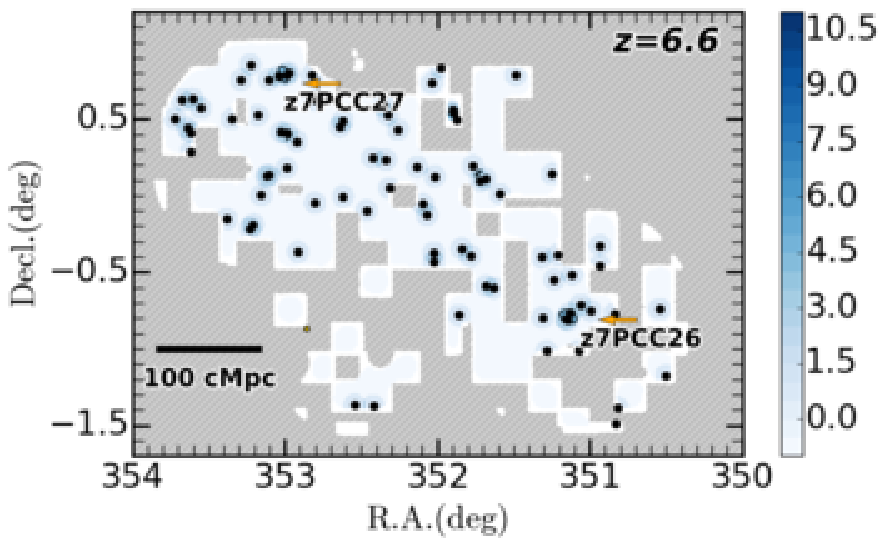}
	\caption{
Same as Figure \ref{fig:coln8uds}, but for the $z=6.6$ LAEs in D-DEEP2-3 field.
	}
	\label{fig:coln9dco}
\end{figure*}


We calculate LAE overdensities in each field with the HSC LAE samples. 
The definition of the LAE overdensity for our observational data
is the same as the one for the model shown in Equation \ref{eq:delta}. 
We use a cylinder with a radius of 0.07 $\rm deg$ corresponding to 10 cMpc at $z \sim 6$. 
The height of the cylinder along a line of sight is 40 cMpc same as the width of 
the redshift distribution of the HSC LAEs. 

Because some regions of the HSC narrowband data are not deep enough
to calculate $\delta$ due to the data quality, we should not use the
HSC LAEs found in the shallow regions for the density evaluation.
The HSC imaging data are divided into $1.7\times1.7\: \rm deg^2$ rectangular tracts
that are made of $0.2 \times 0.2\: \rm deg^2$ rectangular patches. 
We estimate a $5\sigma$ limiting magnitude of each patch in the $NB816$ ($NB921$) data
for $z=5.7$ (6.6) LAEs. We evaluate $\delta$ only in an area where the $5\sigma$ limiting 
magnitude of the $NB816$ ($NB921$) band is brighter than 24.5 (25.0) mag. 
These magnitude limits are determined to keep a high-detection completeness of LAEs \citep{konno2017}. 
We assume that the number density of LAEs in the masked regions is the same as 
the mean number density of LAEs in all fields. 
We also do not evaluate $\delta$ for a cylinder, in which more than $50 \%$ of the area is masked. 
We show the HSC LAE sky distribution and the overdensity maps at $z=5.7$ and $6.6$ 
in Figures \ref{fig:coln8uds}-\ref {fig:coln9dco}. 
The solid lines correspond to contours of $\delta$ from 5 (3) $\sigma$ to 8 (7) $\sigma$ significance levels 
with a step of $1 \sigma$ at $z=5.7$ ($6.6$). Note that the peak of the overdensity is not 
always centered at the highest density region. 
This is because the position of the peak has an uncertainty on a scale of 0.07 deg. 

\subsubsection{Overdensity Identifications}\label{sec:ovid}
We find that $\delta$ values of the HSC LAEs in some regions 
significantly exceed beyond 
those expected by random distribution. 
These $\delta$ values are not explained by a random distribution of galaxies, 
but physical structures. We define a region with $\delta$ exceeding the $5 \sigma$ level
of the Poisson distribution as a high-density region (HDR). At $z=5.7$ $(6.6)$, 
$\delta=9.7$ $(6.6)$ corresponds to the $\simeq 5 \sigma$ significance level.
We find 14 (27) HDRs at $z=5.7$ ($6.6$) with $\delta>9.7$ $(6.6)$. 
There is an overdensity of $z=6.6$ LAEs at R.A.$=34.64$ deg and decl. $= -4.56$
whose $\delta$ is 6.1 slightly below the $\simeq 5 \sigma$ significance level.
This overdensity is reported by \cite{chanchaiworawit2017}. Although this 
does not meet the criterion of $\delta > 6.6$, we include this overdensity 
to the sample of our HDRs. We thus obtain 14 (28) HDRs at $z=5.7$ ($6.6$).

\begin{figure}[!b]
	\plotone{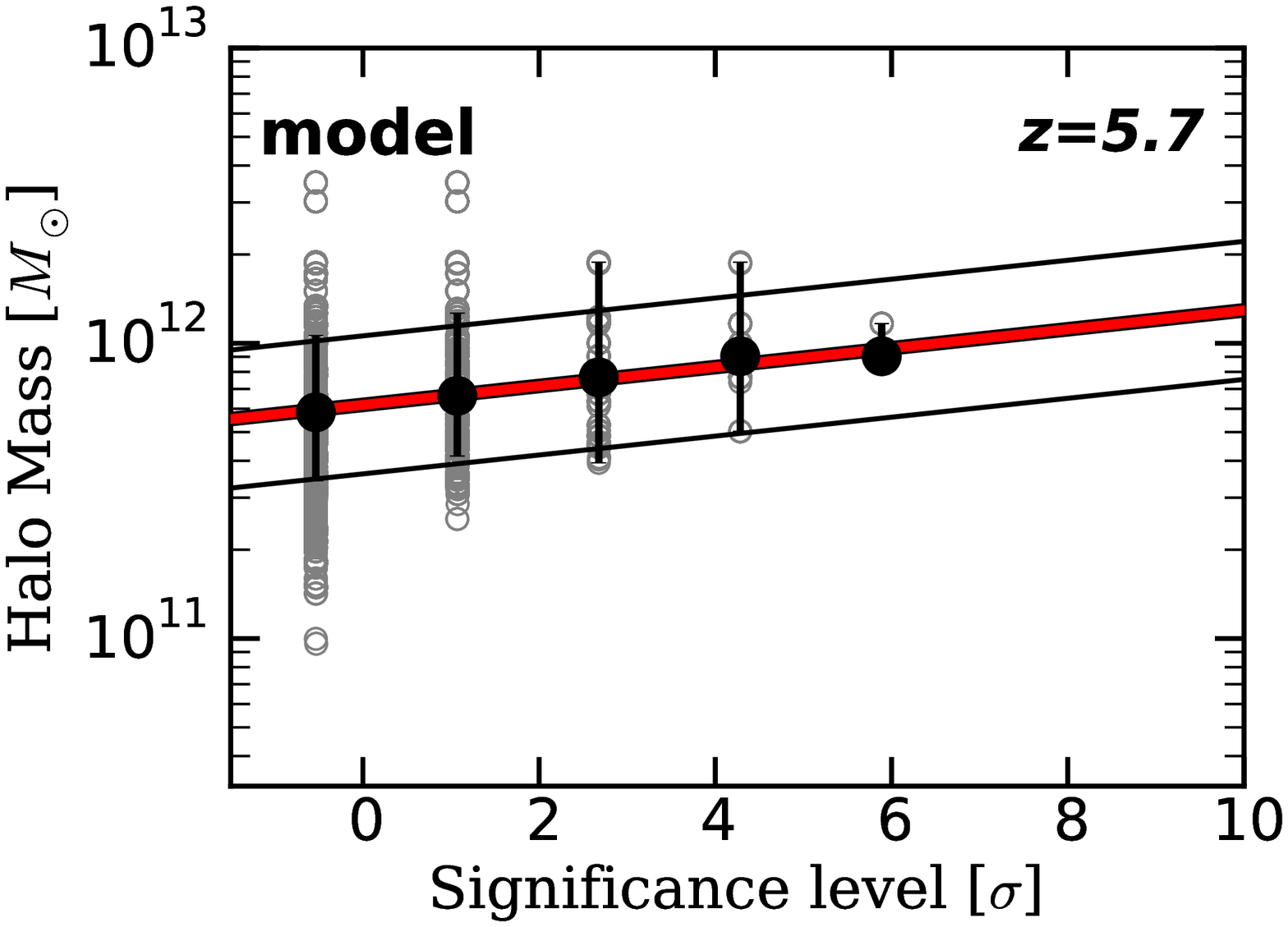}
	\label{fig: mdeln8}
	\caption{
Halo mass as a function of overdensity $\delta$ significance level 
for the model LAEs at $z=5.7$ (gray open circles) produced in the model of \cite{inoue2017}. 
The black filled circles with the error bars indicate the median values of the model LAEs.
The red and black lines represent the best-fit linear function and the 68 $\%$ distribution
of the model LAEs.
	}
	\label{fig: mdeln8}
\end{figure}

\begin{figure}
	\plotone{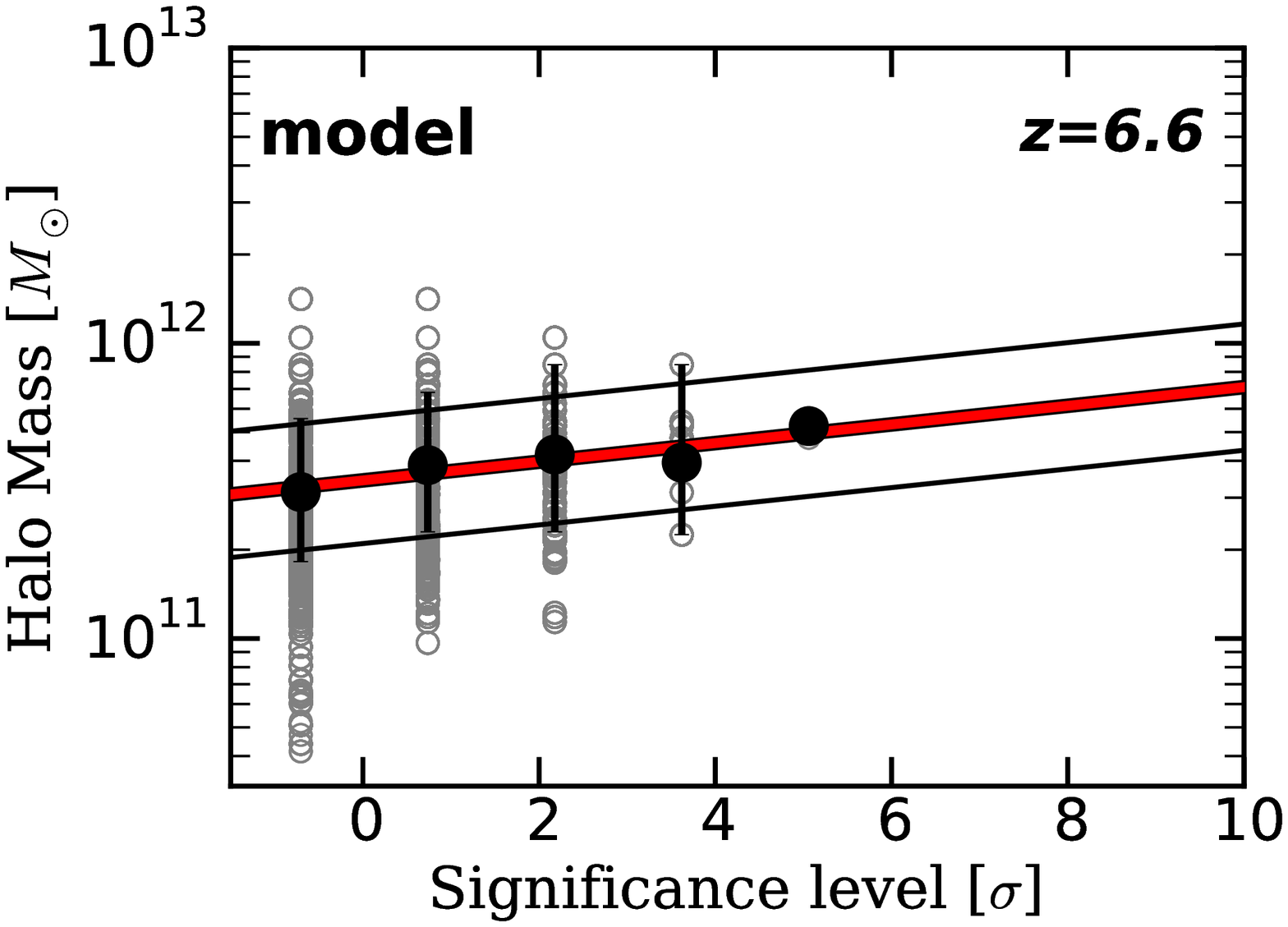}
	\caption{
Same as Figure \ref{fig: mdeln8}, but for the $z=6.6$ model LAEs. 
	}
	\label{fig: mdeln9}
\end{figure}

\subsubsection{Halo Mass Estimates}
From the theoretical model of \cite{inoue2017},
we obtain the halo mass $M_{\rm h}$
as a function of overdensity $\delta$. 
Because the halo mass is strongly related with the structure formation
tightly connected with the abundance of halos and galaxies,
we use LAEs in the model of \cite{inoue2017} 
whose abundance is the same as those of the HSC LAEs. 
We define $M_{\rm h}$ as the most massive halo found in a cylinder volume
used for the $\delta$ calculation.
Figure \ref {fig: mdeln8} (\ref {fig: mdeln9}) shows $M_{\rm h}$ as a function of $\delta$ significance level
at $z=5.7\: (6.6)$. 
We fit the $M_{\rm h}$ - $\delta$ relation with a linear function, and obtain 
$\log_{10}[M_{\rm h}/M_{\odot}]=0.032 \delta + 11.79$ $(\log_{10}[M_{\rm h}/M_{\odot}]= 0.032\delta + 11.54)$ at $z=5.7$ $(6.6)$. 

We use the extended Press-Schechter model of \cite{hamana2006} 
to estimate the present-day halo masses of the high-$z$ ($z=5.7$ and $6.6$) halos. 
Based on the $M_{\rm h}$ - $\delta$ relation,
we find that $\>$60 (58)\% of the $z=5.7$ ($6.6$) $M_{\rm h}$-halos in the HDRs 
are expected to evolve into present-day cluster haloes with a mass of $>10^{14}\:\rm M_{\odot}$ by $z=0$. 
Because more than a half of the $M_{\rm h}$-halos in the HDRs are progenitors of the present-day clusters, 
we regard the 14 (28) HDRs at $z=5.7$ ($6.6$) as protocluster candidates. 
The 14 (28) protocluster candidates are listed in Table \ref{tab: pcclist}. 
Here we name the $z=5.7$ (6.6) protocluster candidates as HSC-z6 (7) PCC. 

We compare the abundance of the protocluster candidates with that of present-day clusters. 
The comoving survey volumes of the HSC observations are $\sim1.2\times10^{7}\:\rm Mpc^3$ and 
$\sim 1.9\times 10^{7}\:\rm Mpc^3$ at $z=5.7$ and 6.6, respectively. 
Because there exists one present-day cluster with a mass of $1-3\times 10^{14} M_\odot$ 
in a volume of $\sim 5\times 10^5$ Mpc$^3$ \citep{reiprich2002}, 
it is expected that our survey volumes at $z=5.7$ and $6.6$
include $\sim20$ and $\sim 40$ present-day clusters, respectively.
These numbers are comparable with those of our protocluster candidates, 14 and 28. 

\subsubsection{Three-Dimensional Distribution and Protocluster Candidates}\label{sec:three}
\begin{figure}
	\plotone{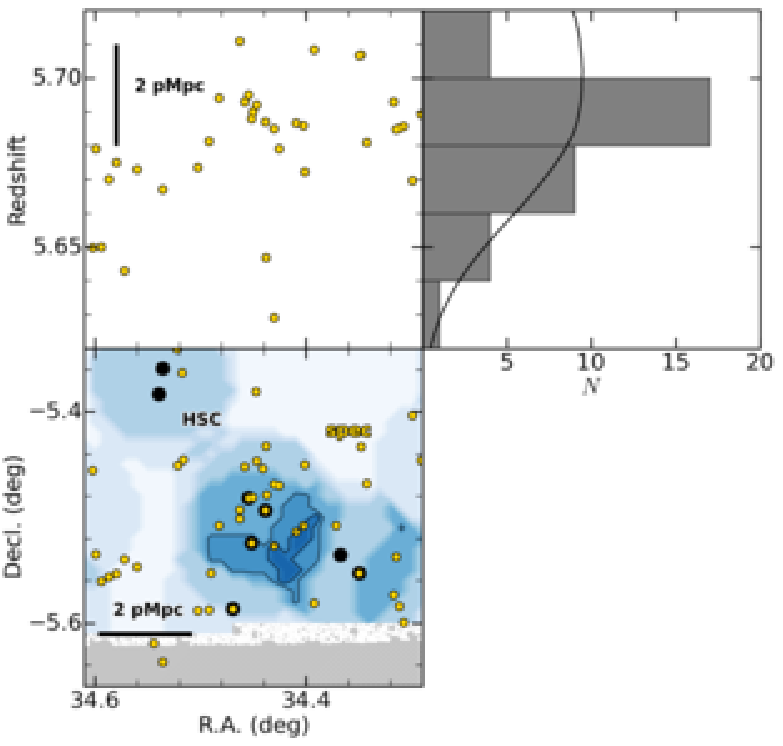}
	\caption{
Three-dimensional map of HSC-z6PCC1. 
The bottom panel presents the distribution of the LAEs projected on the sky. 
The top-left panel shows the distribution of the LAEs on the plane of 
transverse (east to west) vs. radial (redshift) directions. 
The black filled circles represent the HSC LAEs with $NB816<24.5$
used for the overdensity evaluation, while the yellow filled circles denote 
the spectroscopically-confirmed LAEs that include faint sources with $NB816\simeq 25-26$.
The black solid lines in the bottom panel indicate 
the contours of the overdensity significance levels from 5$\sigma$ to 8$\sigma$ with a step of 1$\sigma$. 
The masked regions are shown with the gray regions. 
The top right panel shows the redshift distribution of the spectroscopically-confirmed LAEs. 
The black line indicates the mean expected number of LAEs in the region.
}
	\label{fig:hscz6pcc1}
\end{figure}

\begin{figure}
	\plotone{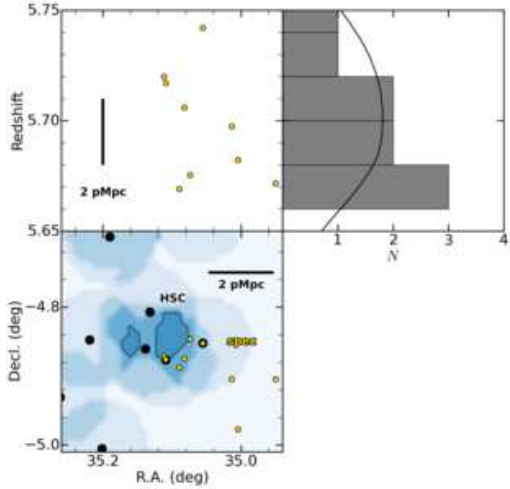}
	\caption{
Same as Figure \ref{fig:hscz6pcc1}, but for HSC-z6PCC4. 
	}
	\label{fig:hscz6pcc2}
\end{figure}

\begin{figure}
	\plotone{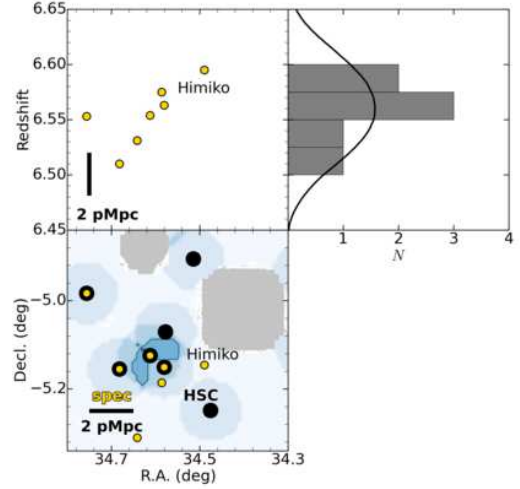}
	\caption{
Same as Figure \ref{fig:hscz6pcc1}, but for HSC-z7PCC9. 
The black filled circles represent the HSC LAEs with $NB921<25.0$.
The yellow circles indicate the spectroscopically-confirmed LAEs 
including sources with $NB921\simeq 25-26$. 
	}
	\label{fig:hscz7pcc1}
\end{figure}

\begin{figure}
	\plotone{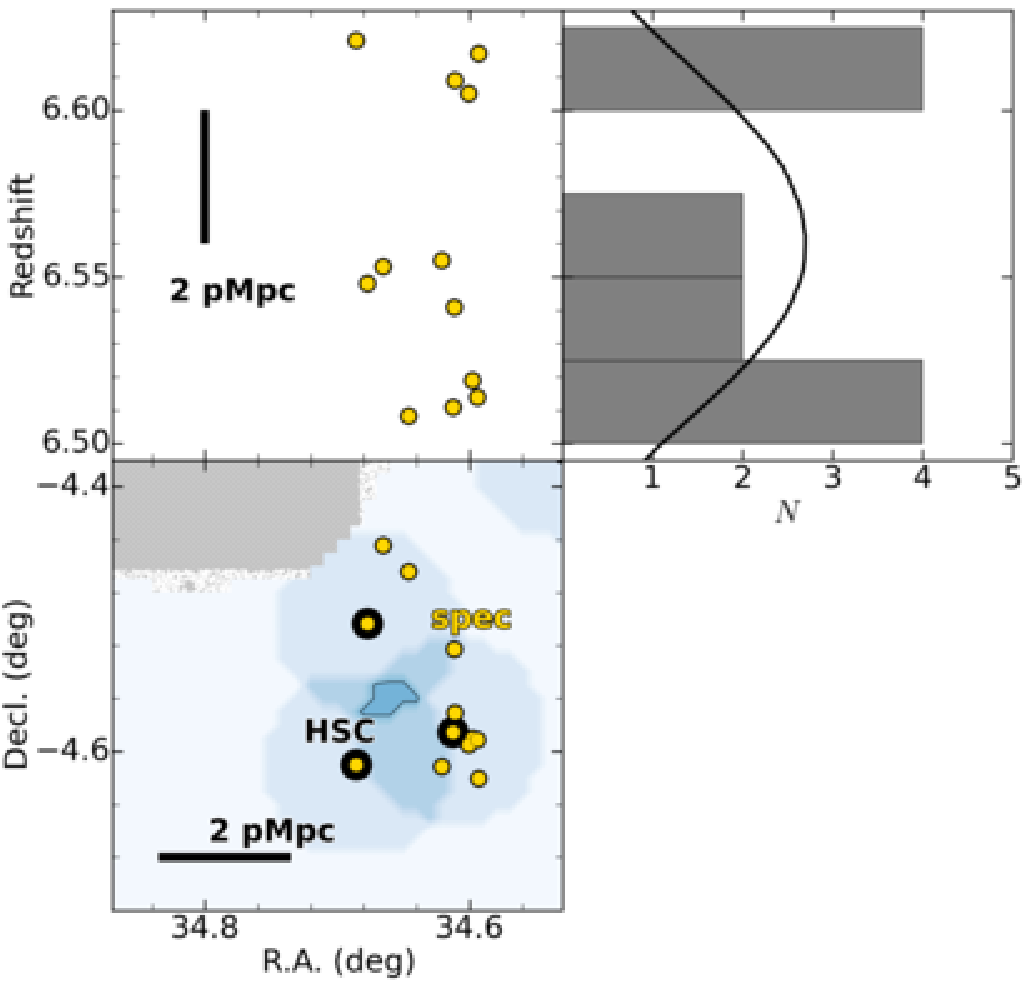}
	\caption{
Same as Figure \ref{fig:hscz7pcc1}, but for HSC-z7PCC28. 
	}
	\label{fig:hscz7pcc2}
\end{figure}

Based on the follow-up spectroscopic observations in Section \ref{sec:spec}, we find 3 (3) protocluster candidates at $z=5.7 \: (6.6)$ which have (a) spectroscopically confirmed LAE(s). 
These are HSC-z6PCC1, HSC-z6PCC4, and HSC-z6PCC5 (HSC-z7PCC3, HSC-z7PCC9, and HSC-z7PCC28)
at $z=5.7$ ($6.6$). The three-dimensional distributions of HSC-z6PCC1, HSC-z6PCC4, 
HSC-z7PCC9, and HSC-z7PCC28 are shown in Figures \ref{fig:hscz6pcc1},
\ref{fig:hscz6pcc2}, \ref{fig:hscz7pcc1}, \ref {fig:hscz7pcc2}, respectively. 
Here we explain three examples of the protocluster candidates, 
HSC-z6PCC1, HSC-z7PCC9, and HSC-z7PCC28. 
\subparagraph{HSC-z6PCC1} 
HSC-z6PCC1 (Figure \ref {fig:hscz6pcc1}) consists of $z=5.7$ LAEs in the southern part of UD SXDS. 
Twelve spectroscopically confirmed LAEs exist within a distance of $\sim 1$ physical Mpc (pMpc). 
The redshift averaged over the spectroscopically-confirmed LAEs is $z=5.692$.
HSC-z6PCC1 is the same structure as Clump A that is a protocluster identified by \cite{ouchi2005}. 
Six out of the 12 spectroscopically confirmed LAEs are included in Clump A. 
\subparagraph{HSC-z7PCC9} 
HSC-z7PCC9 at $z=6.6$ (Figure \ref {fig:hscz7pcc1}) is located at the center of UD SXDS. 
HSC-z7PCC9 consists of five spectroscopically confirmed LAEs, including the giant Ly$\alpha$ nebula 'Himiko' \citep{ouchi2009}. 
The average redshift of the LAEs is $z=6.574$.
If all of the LAEs of HSC-z7PCC9 are spectroscopically confirmed, HSC-z7PCC9 could be
one of the earliest protoclusters found to date.
\subparagraph{HSC-z7PCC28} 
HSC-z7PCC28 (Figure \ref {fig:hscz7pcc2}) is placed at the northern part of UD SXDS at $z=6.6$. 
This is the protocluster candidate reported by \cite{chanchaiworawit2017},
although the overdensity of HSC-z7PCC28 is $\delta=6.1$ slightly below
the $5 \sigma$ significance level (Section \ref{sec:ovid}).
There are five spectroscopically confirmed LAEs in a sphere with a radius of $\sim1$ pMpc.
The redshift averaged over the spectroscopically-confirmed LAEs is $z=6.534$. 
Three out of the five spectroscopically confirmed LAEs are included 
in the members of the overdensity shown in \cite{chanchaiworawit2017}.  

\subsection{Implications for Cosmic Reionization}\label{sec:imp}
\subsubsection{Spatial Correlation between Bright LAEs and Overdensities}
\begin{figure}
	\plotone{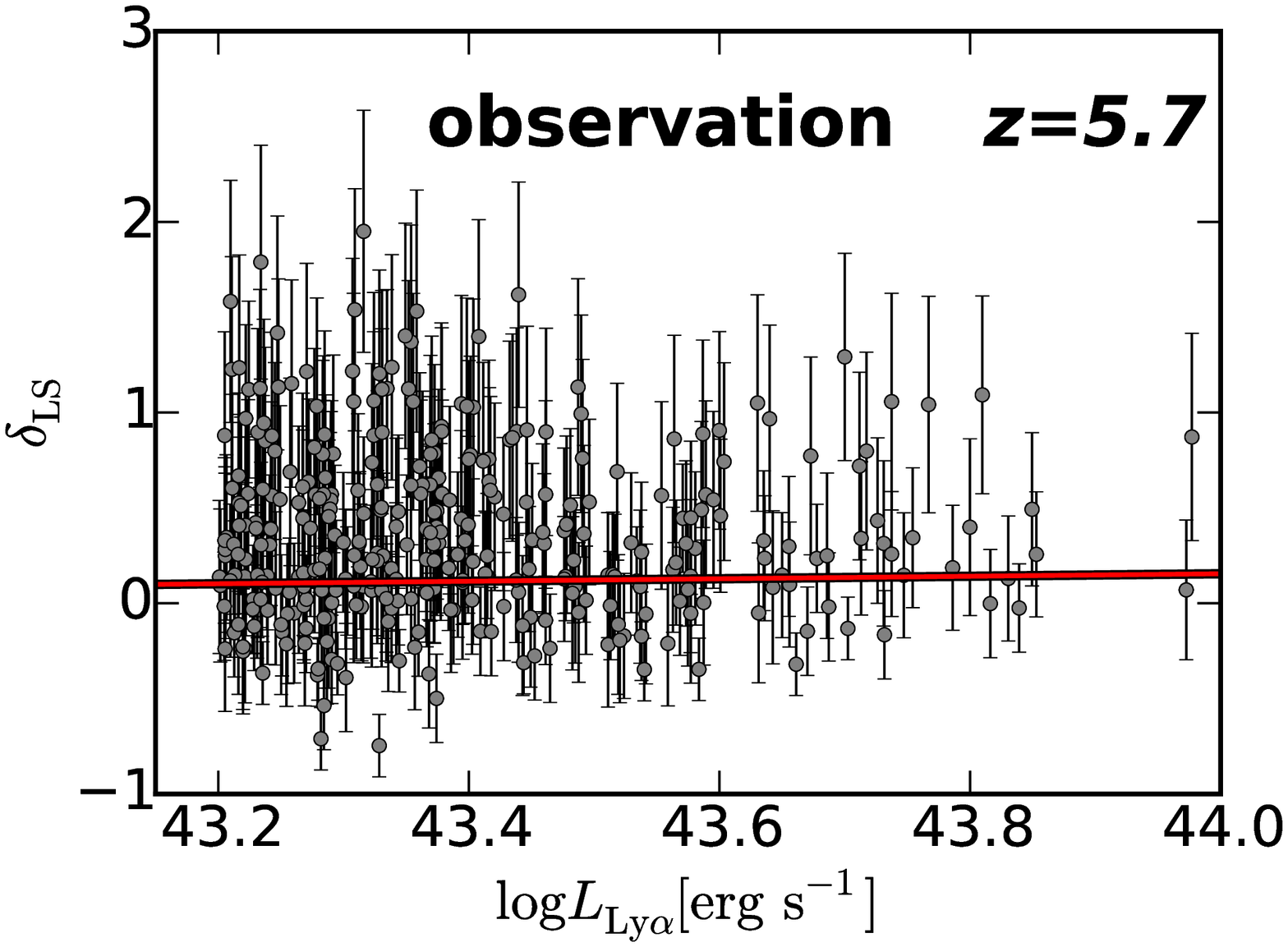}
	\caption{
Ly$\alpha$ luminosity $L_{\rm Ly \alpha}$ as a function of
large-scale overdensity $\delta_{\rm LS}$ for the HSC LAEs at $z=5.7$ (gray circles). 
The red line indicates the best-fit linear function. 
	}
	\label{fig:lumideln8}
\end{figure}

\begin{figure}
	\plotone{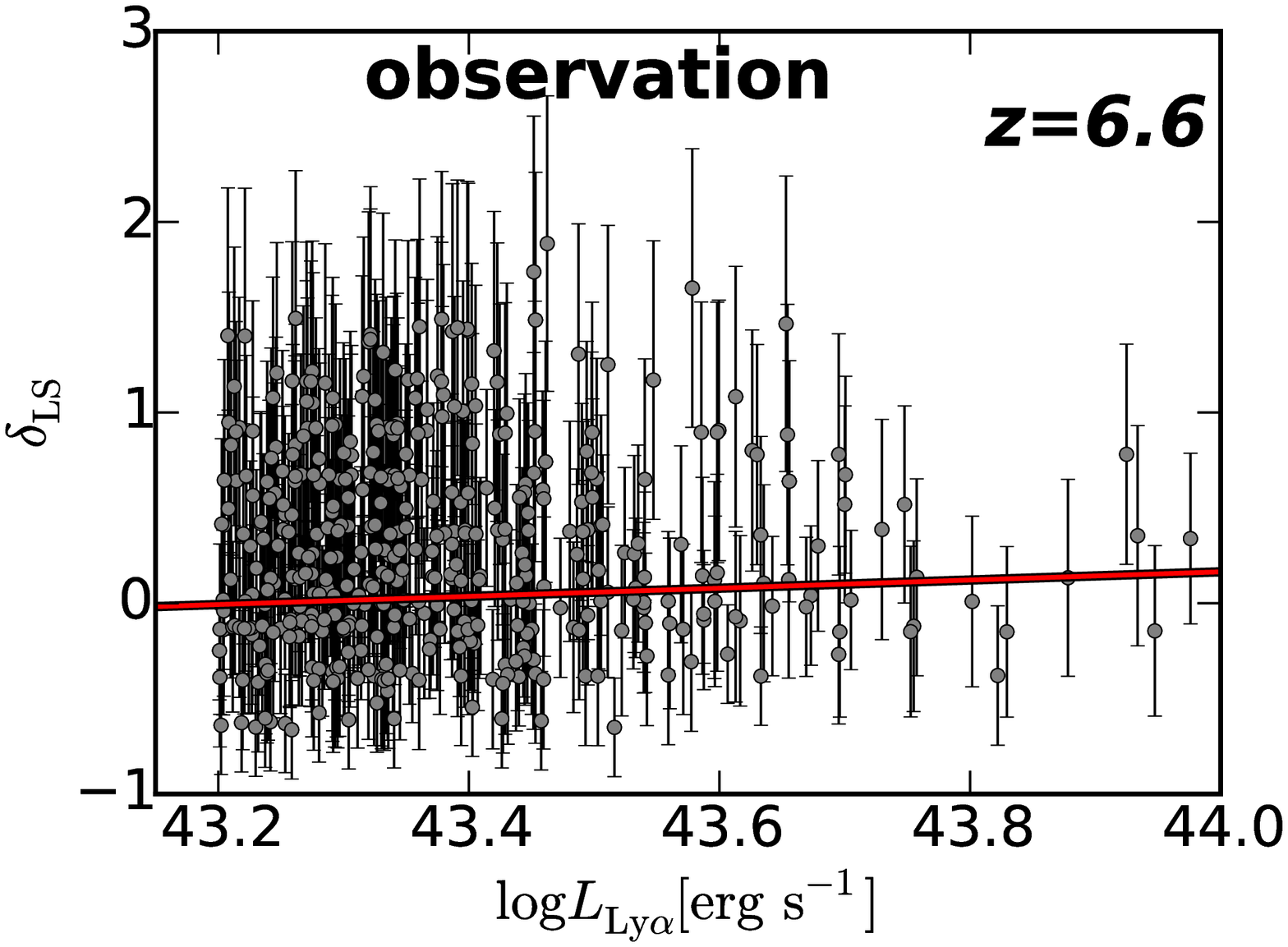}
	\caption{
Same as Figure \ref{fig:lumideln8}, but for the HSC LAEs at $z=6.6$.
	}
	\label{fig:lumideln9}
\end{figure}

To study the origin of the bright-end excess of Ly$\alpha$ luminosity functions at $z=5.7$ and $6.6$ \citep{konno2017}, we investigate the correlation between Ly$\alpha$ luminosity and overdensity. 
Figure \ref {fig:lumideln8} (\ref {fig:lumideln9}) shows the relation between Ly$\alpha$ luminosity $L_{\rm Ly \alpha}$
and large-scale LAE overdensity $\delta_{\rm LS}$ for $z=5.7$ ($6.6$) LAEs. 
Here $\delta_{\rm LS}$ is defined with a circle with a radius of 0.20 $\rm deg$ that corresponds
to $\sim30$ cMpc at $z \sim 6$ comparable with the size of typical ionized bubbles at this redshift
predicted by \cite{furlanetto2006} (cf. $\delta$ defined with a circular radius of 0.07 deg; see Section \ref{sec:spa}).
With the results of Figures  \ref {fig:lumideln8} and \ref {fig:lumideln9},
we calculate a Spearman's rank correlation coefficient $\rho$ and a p-value to test the existence of 
the correlation between $L_{\rm Ly \alpha}$ and $\delta_{\rm LS}$. 
We obtain $\rho=-0.017$ $(0.020)$ with p-value$=0.75$ $(0.68)$ for $z=5.7$ ($6.6$) LAEs, which suggest that there are no significant correlations between $L_{\rm Ly \alpha}$ and $\delta_{\rm LS}$. 
This result indicates that bright $L_{\rm Ly \alpha}$ LAEs are not selectively
placed at the overdensity and that there is no clear evidence connecting
the bright-end LF excess and the ionized bubble. 
Because the statistical uncertainty of this analysis is still large,
it is not a conclusive result. However, there is an increasing
possibility that the ionized bubbles and the bright-end LF excess
may not be related.
For the other possible origins of the bright-end excess, \cite{konno2017} 
discuss the AGN/low-$z$ contamination and the blended merging galaxies. 
We should discuss these other possibilities more seriously.

\subsubsection{Correlation between Ly$\alpha \ $ EW and Overdensity} \label {sec: ewdel}
\begin{figure}
	\plotone{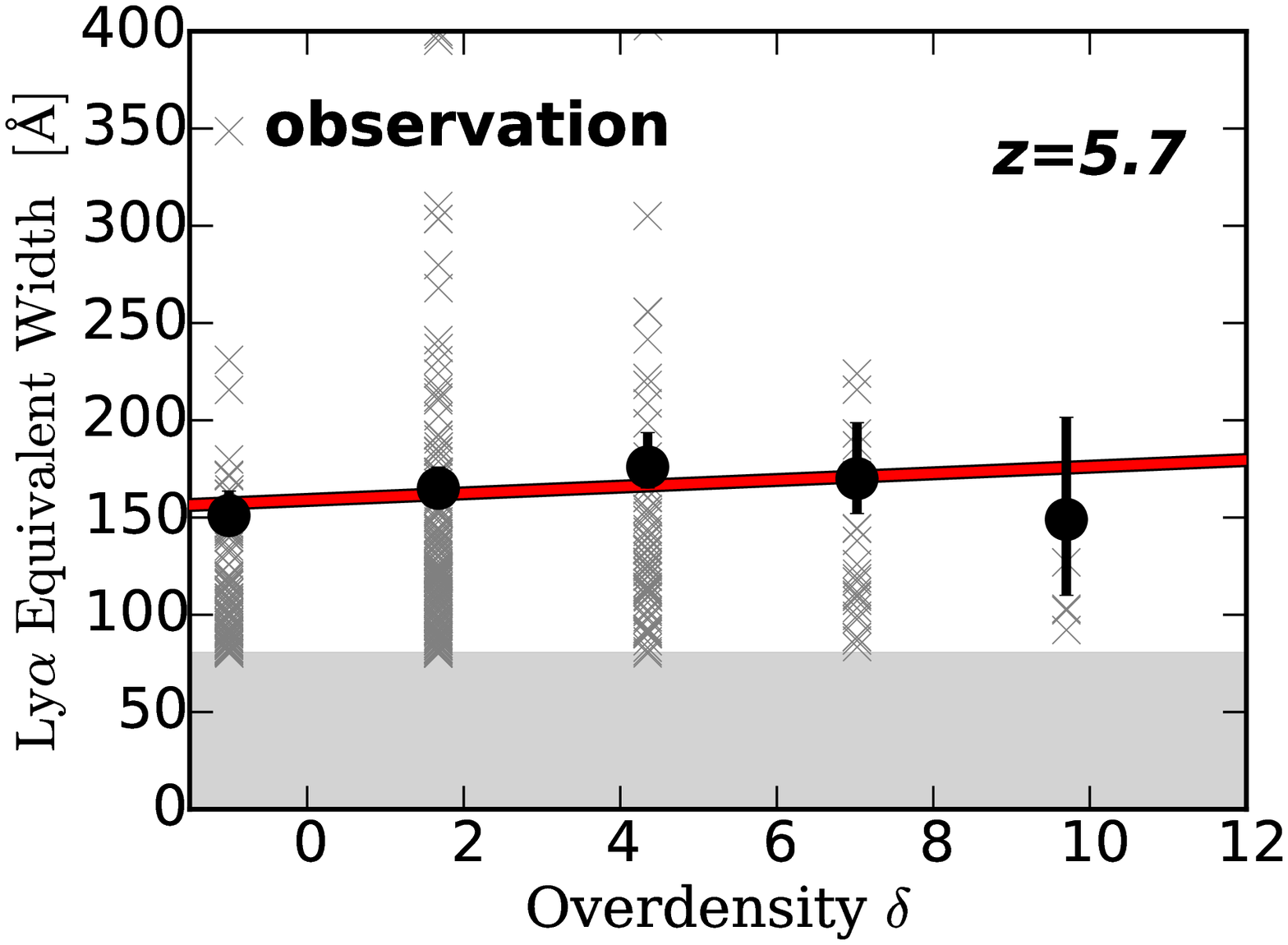}
	\caption{
Ly$\alpha$ EW and overdensity $\delta$ for the HSC LAEs at $z=5.7$ (gray crosses). 
The black circles with the error bars indicate the median values of the HSC LAEs
at a given $\delta$. The red line represents the best-fit linear function.
The gray region indicates the Ly$\alpha$ EW selection limit.
	}
	\label{fig:ewdeln8}
\end{figure}

\begin{figure}
	\plotone{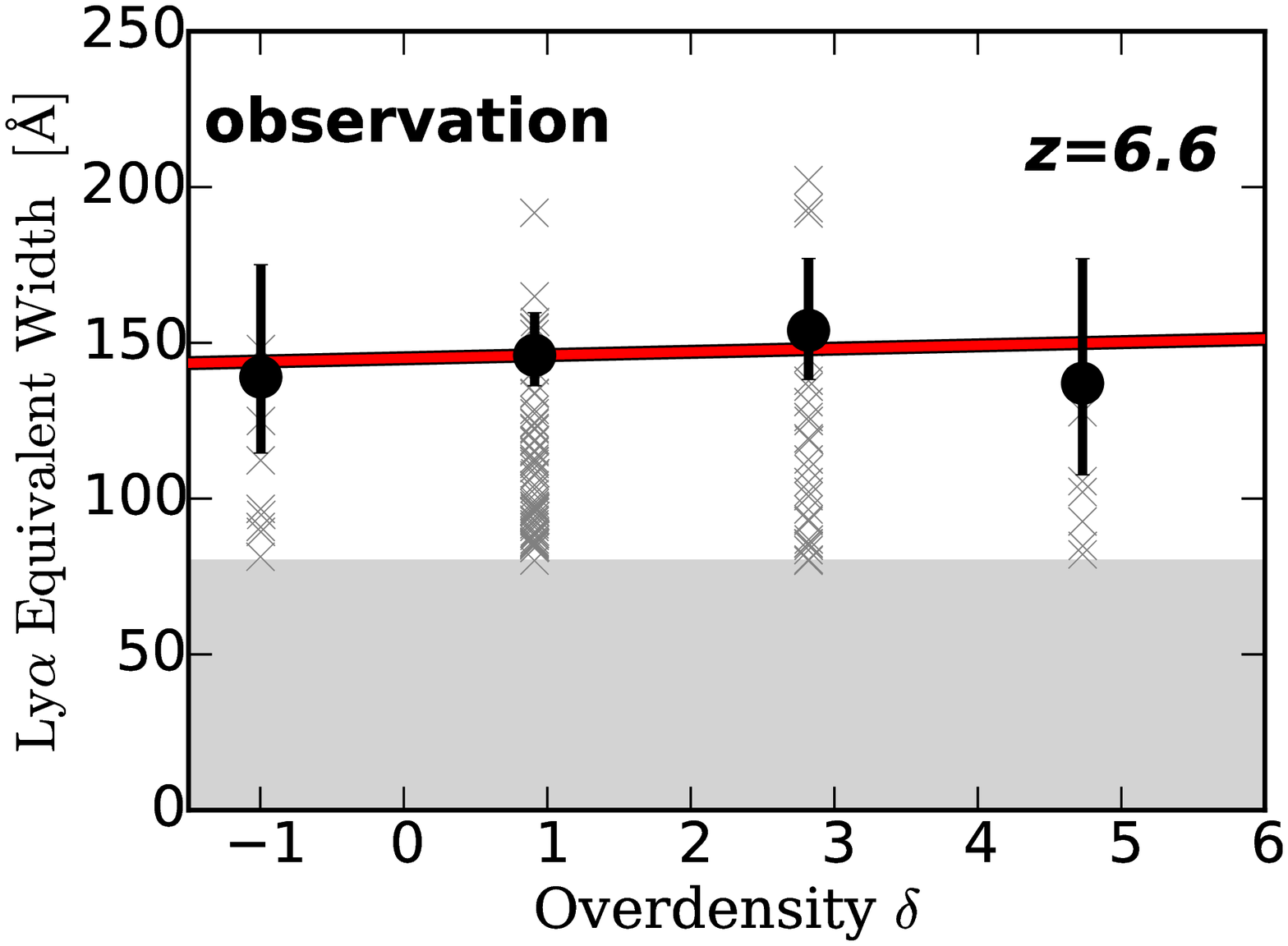}
	\caption{
Same as Figure \ref{fig:ewdeln8}, but for the HSC LAEs at $z=6.6$.
	}
	\label{fig:ewdeln9}
\end{figure}

Figure \ref {fig:ewdeln8} (\ref {fig:ewdeln9}) presents $EW^{\rm rest}_{\rm Ly\alpha}$ as a function of $\delta$ at $z=5.7$ $(6.6)$. 
$EW^{\rm rest}_{\rm Ly\alpha}$ is estimated in the same manner as \cite{shibuya2017a}. 
We calculate $EW^{\rm rest}_{\rm Ly\alpha}$ of LAEs from the $NB816$ ($NB921$) and $z$ ($y$) band magnitudes. 
We use the subsamples of the LAEs in a range of $\delta$, and obtain
a median value of $EW^{\rm rest}_{\rm Ly\alpha}$ at a given $\delta$.
We perform chi-square fitting of the linear function 
to the $EW^{\rm rest}_{\rm Ly\alpha}$ - $\delta$ relations, 
and obtain the best-fit parameters, $\alpha$ and $EW_{\delta=0}$, defined in Section \ref{sec:theo}. 
Because the theoretical model predicts that
the value of $\alpha$ increases from $z=5.7$ to $6.6$ (Section \ref{sec:theo}),
we show redshift evolution of $\alpha$ of the observational results in Figure \ref {fig: sigcont_sim}. 
Figure \ref {fig: sigcont_sim} indicates that there is no significant evolution of the $EW^{\rm rest}_{\rm Ly\alpha}$ - $\delta$ relation from $z=5.7$ to $6.6$ beyond the uncertainties accomplished with our HSC data so far obtained. 

\subsubsection{Comparison with the Theoretical Model} \label{sec: model}
\begin{figure}
	\plotone{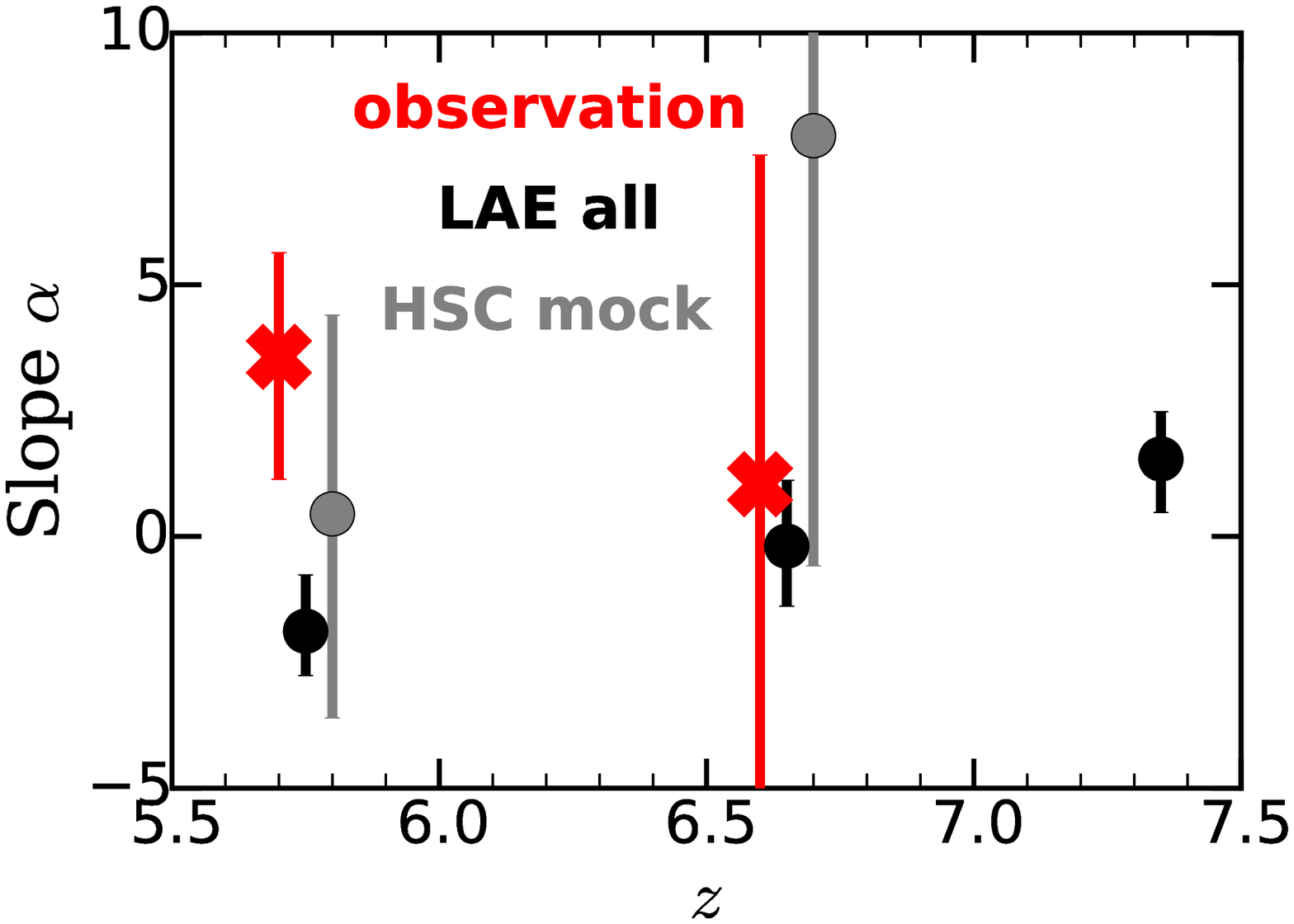}
	\caption{
Redshift evolution of the slope $\alpha$. 
The red crosses denote the HSC LAEs, while  
the gray (black) circles show the model predictions with the samples of 'HSC mock' and LAE all', respectively. 
To avoid overlaps of the symbols, we slightly shift black and grey circles by 0.05 and 0.10 in redshift, respectively. 
	}
	\label{fig: sigcont_sim}
\end{figure}

We compare the results of Section \ref {sec: ewdel} with the theoretical model of \cite{inoue2017}. 
We select mock LAEs which are brighter than 25.0 mag in narrowbands,
which is the same magnitude limit as the HSC LAE $\delta$ estimates.
We also apply the selection limits of the Ly$\alpha$ EW which are similar to those of the HSC LAE samples.
We thus obtain 447 (80) mock LAEs for $z=5.7$ (6.6) that are referred to as 'HSC mock'. 
We derive the best-fit parameters and errors for HSC mock 
in the same way as Section \ref {sec: ewdel}. 
Note that we define the error of $EW^{\rm rest}_{\rm Ly\alpha}$ as the range of 68$\%$ distribution. 
Figure \ref {fig: sigcont_sim} presents redshift evolution of the slope $\alpha$ of 'HSC mock' 
at $z=5.7$ and $6.6$. (In this model, the average neutral hydrogen fraction in the IGM at $z=5.7$ and $6.6$ 
are $\log{x_{\rm HI}}=-3.9$ and $-0.36$, respectively.)
The model does not show the significant evolution of the $EW^{\rm rest}_{\rm Ly\alpha}$ - $\delta$ relation 
beyond the statistical errors, which is consistent with those of the HSC LAE samples.
The model suggests that the present HSC LAE samples are
not large enough to test the existence of the ionized bubbles and the inside-out scenario of 
cosmic reionization. The HSC survey is underway, which will significantly
enlarge the sample with the wider and deeper data for LAEs at $z=5.7$ and $6.6$
and make a new sample of LAEs at $z=7.3$. 
There is a possibility that the evolution of the $EW^{\rm rest}_{\rm Ly\alpha}$ - $\delta$ relation 
from $z=5.7$ to $7.3$ may be identified by the upcoming HSC observations 
providing the large samples of LAEs at $z=5.7-7.3$.
The ionized bubbles and the inside-out scenario 
should be tested in the forthcoming studies with
the large samples of LAEs at $z=5.7-7.3$.

\section{Summary}\label{sec:Summary}

In this study, we study LAE overdensities at $z = 5.7$ and $6.6$ with the early datasets of the HSC SSP survey
based on the 2,230 LAEs obtained in the SILVERRUSH program. 
We identify the LAE overdensities and discuss cosmic reionization with
the properties of LAEs, overdensity $\delta$, Ly$\alpha$ luminosity $L_{\rm Ly \alpha}$, and 
the rest-frame Ly$\alpha$ equivalent width $EW^{\rm rest}_{\rm Ly\alpha}$. 
Our major results are listed below: 
\begin{enumerate}
	\item We calculate the LAE overdensity $\delta$ with the samples of the HSC LAEs at $z=5.7$ and $6.6$. 
	We identify 14 (28) $z=5.7$ $(6.6)$ LAE overdensities with the $\gtrsim 5\sigma$ significance level,
	six out of which have $1-12$ spectroscopically confirmed LAEs.
	We compare the LAE overdensities with the
	the cosmological Ly $\alpha$ radiative transfer models,
	and find that more than a half of these LAE overdensities 
	(60\% and 58\% of the LAE overdensities at $z=5.7$ and $6.6$) are 
	progenitors of the present-day clusters with a mass of $\gtrsim 10^{14} M_\odot$.
	These 14 (28) LAE overdensities are thus protocluster candidates at $z=5.7$ $(6.6)$
	that are listed in Table \ref{tab: pcclist}.
	\item 
	We investigate the correlation between $L_{\rm Ly \alpha}$ and $\delta$ with the HSC LAEs.
	We obtain a Spearman's rank correlation coefficient $\rho=-0.017$ $(0.020)$ with p-value=0.75 $(0.68)$ for 
	$z=5.7$ ($6.6$) LAEs, which indicate that there is no evidence of significant correlations between 
	$L_{\rm Ly \alpha}$ and $\delta$ beyond the observational uncertainties. 
	Our result is related to the recent discussion about the bright-end excess of Ly$\alpha$ LFs at $z=5.7$ and $6.6$ 
	such found in \citet{konno2017}. For the physical reason of the bright-end excess,	
	there is an idea that bright galaxies selectively existing in an overdensity region
	are placed near the center of the ionized bubbles
	that allow Ly$\alpha$ photons escape from the partly neutral IGM at the EoR.
	Because our results show no correlation between $L_{\rm Ly \alpha}$ and $\delta$,
	there is no evidence supporting this idea.	
	\item We study the relations between $EW^{\rm rest}_{\rm Ly\alpha}$ and $\delta$ at $z=5.7$ and $6.6$.
	We fit a linear function to the $EW^{\rm rest}_{\rm Ly\alpha}$-$\delta$ data, and find that the slope (the relation)
	does not evolve (is not steepened) from $z=5.7$ to $6.6$ beyond the errors.
	The cosmological reionization model with the Ly$\alpha$ radiative transfer suggests
	that the slope is steepened towards the early EoR with a high neutral hydrogen fraction 
	in the inside-out reionization scenario, 
	because the ionized bubbles around galaxy overdensities ease the escape of Ly$\alpha$ emission 
	from the partly neutral IGM at the EoR.
	Although the model suggests that the statistical accuracy of our HSC data
	is not high enough to investigate this steepening, so far we find no such steepening
	in the available HSC data.
	There is a possibility of detecting the evolution of the 
	$EW^{\rm rest}_{\rm Ly\alpha}$ - $\delta$ relation from $z=5.7$ to $7.3$
	by the scheduled HSC narrowband observations 
	that will make larger samples of LAEs at $z=5.7-6.6$ as well as a new sample of LAEs at $z=7.3$. 
\end{enumerate}

\acknowledgments
We are grateful to 
Richard S. Ellis, 
Ryohei Itoh, 
Shotaro Kikuchihara, 
Haruka Kusakabe, 
Hilmi Miftahul, 
Shiro Mukae, 
Yuma Sugahara, 
Hidenobu Yajima, 
Haibin Zhang, 
and 
Zheng Zheng for useful comments and discussions. 
We thank Shingo Shinogi for providing information about the Keck/DEIMOS spectra of $z=5.7$ LAEs. 
We also thank Janice Lee and Ivelina Momcheva for providing Magellan/IMACS
spectra of $z=5.7$ LAEs that were taken as mask fillers.
The Hyper Suprime-Cam (HSC) collaboration includes the astronomical communities of Japan and Taiwan, and Princeton University.  The HSC instrumentation and software were developed by the National Astronomical Observatory of Japan (NAOJ), the Kavli Institute for the Physics and Mathematics of the Universe (Kavli IPMU), the University of Tokyo, the High Energy Accelerator Research Organization (KEK), the Academia Sinica Institute for Astronomy and Astrophysics in Taiwan (ASIAA), and Princeton University.  Funding was contributed by the FIRST program from Japanese Cabinet Office, the Ministry of Education, Culture, Sports, Science and Technology (MEXT), the Japan Society for the Promotion of Science (JSPS),  Japan Science and Technology Agency  (JST),  the Toray Science  Foundation, NAOJ, Kavli IPMU, KEK, ASIAA,  and Princeton University.
The Pan-STARRS1 Surveys (PS1) have been made possible through contributions of the Institute for Astronomy, the University of Hawaii, the Pan-STARRS Project Office, the Max-Planck Society and its participating institutes, the Max Planck Institute for Astronomy, Heidelberg and the Max Planck Institute for Extraterrestrial Physics, Garching, The Johns Hopkins University, Durham University, the University of Edinburgh, Queen's University Belfast, the Harvard-Smithsonian Center for Astrophysics, the Las Cumbres Observatory Global Telescope Network Incorporated, the National Central University of Taiwan, the Space Telescope Science Institute, the National Aeronautics and Space Administration under Grant No. NNX08AR22G issued through the Planetary Science Division of the NASA Science Mission Directorate, the National Science Foundation under Grant No. AST-1238877, the University of Maryland, and Eotvos Lorand University (ELTE).
This paper makes use of software developed for the Large Synoptic Survey Telescope. We thank the LSST Project for making their code available as free software at http://dm.lsst.org.
Based in part on data collected at the Subaru Telescope and retrieved from the HSC data archive system, which is operated by the Subaru Telescope and Astronomy Data Center at National Astronomical Observatory of Japan.
The NB816 filter was supported by Ehime University (PI: Y. Taniguchi). 
The NB921 filter was supported by KAKENHI (23244025) Grant-in-Aid for Scientific Research (A) through the Japan Society for the Promotion of Science (PI: M. Ouchi). 
This work is supported by World Premier International Research 
Center Initiative (WPI Initiative), MEXT, Japan, and 
KAKENHI (15H02064) Grant-in-Aid for Scientific Research (A) 
through Japan Society for the Promotion of Science.

\begin{deluxetable*}{ccccc}
\tablecolumns{5}
\tablecaption{Spectroscopic Sample of the $z=5.7$ LAEs
\label{tab:spec_n8uds}}
\tablehead{
	\colhead{ID} & \colhead{R.A. (J2000)} & \colhead{Decl. (J2000)} & \colhead{$z$} & \colhead{Reference} \\
	\colhead{(1)} & \colhead{(2)} & \colhead{(3)} & \colhead{(4)} & \colhead{(5)} 
}
\startdata
	\multicolumn{5}{c}{UD SXDS}\\ \hline
HSC J021714-050844  & $34.3104$  &  $-5.1456$  &  $5.685$  &  This Study\\
HSC J021712-050748  & $34.3027$  &  $-5.1302$  &  $5.699$  &  This Study\\
HSC J021728-051217  & $34.3678$  &  $-5.2047$  &  $5.676$  &  This Study\\
HSC J021750-050203  & $34.4619$  &  $-5.0342$  &  $5.708$  &  This Study\\
... & ... & ... & ... & ...
\enddata
\tablecomments{
Spectroscopically identified LAEs at $z=5.7$. See the full sample catalog in the version published in ApJ. (1) Object ID; (2) right ascension; (3) declination; (4) spectroscopic redshift; (5) reference of the spectroscopic redshift. O05 = \cite{ouchi2005}, O08 = \cite{ouchi2008}, M12 = \cite{mallery2012}, and SH17 = \cite{shibuya2017b}. 
}
\end{deluxetable*}

\begin{deluxetable*}{ccccc}
\tablecolumns{5}
\tablecaption{Spectroscopic Sample of the $z=6.6$ LAEs
\label{tab:spec_n9uds}}
\tablehead{
	\colhead{ID} & \colhead{R.A. (J2000)} & \colhead{Decl. (J2000)} & \colhead{$z$} & \colhead{Reference} \\
	\colhead{(1)} & \colhead{(2)} & \colhead{(3)} & \colhead{(4)} & \colhead{(5)} 
	}
\startdata
	\multicolumn{5}{c}{UD SXDS}\\ \hline
HSC J021703-045619  & $34.2644$  &  $-4.9386$  &  $6.589$  &  O10\\
HSC J021820-051109  & $34.5862$  &  $-5.1861$  &  $6.575$  &  O10\\
HSC J021819-050900  & $34.5808$  &  $-5.1502$  &  $6.563$  &  O10\\
HSC J021757-050844  & $34.4899$  &  $-5.1457$  &  $6.595$  &  O10\\
... & ... & ... & ... & ...
\enddata
\tablecomments{
Spectroscopically identified LAEs at $z=6.6$. See the full sample catalog in the version published in ApJ. (1) Object ID; (2) right ascension; (3) declination; (4) spectroscopic redshift; 
(5) reference of the spectroscopic redshift. O10 = \cite{ouchi2010}, SO15 = \cite{sobral2015}, HU16 = \cite{hu2016}, SH17 = \cite{shibuya2017b}, and C$\&$G17 = \cite{chanchaiworawit2017} and \cite{guzman2017}. 
}
\end{deluxetable*}

\begin{deluxetable*}{lccclcclll}
\tablecaption{Protocluster Candidates
}
\tablehead{
\colhead{Name}&
\colhead{Layer}&
\colhead{Field} & 
\colhead{R.A. (J2000)}&
\colhead{Decl. (J2000)}&
\colhead{overdensity $\delta$}&
\colhead{significance}&
\colhead{$n_{\rm photo}$}&
\colhead{$n_{\rm spec}$}&
\colhead{$z_{\rm spec}$}\\
\colhead{(1)}&
\colhead{(2)}&
\colhead{(3)}&
\colhead{(4)}& 
\colhead{(5)}&
\colhead{(6)}&
\colhead{(7)}&
\colhead{(8)}&
\colhead{(9)}&
\colhead{(10)}
}
\startdata
	\multicolumn{10}{c}{$z=5.7$}\\ \hline
HSC-z6PCC3 & UD & SXDS & 34.26 & $-4.32$ & 9.7 & 5.4 & 4 (5) & 0 & - \\
HSC-z6PCC1 & UD & SXDS & 34.42 & $-5.54$ & 15.0 & 8.4 & 6 (7) & 12 & 5.692 \\
HSC-z6PCC4 & UD & SXDS & 35.16 & $-4.85$ & 9.7 & 5.4 & 4 (7) & 4 & 5.719 \\

HSC-z6PCC5 & UD & COSMOS & 149.94 & 1.60 & 9.7 & 5.4 & 4 (5) & 2 & 5.686 \\

HSC-z6PCC6 & Deep & ELAIS-N1 & 241.84 & 54.27 & 9.7 & 5.4 & 4 (4) & 0 & - \\
HSC-z6PCC7 & Deep & ELAIS-N1 & 242.32 & 53.77 & 9.7 & 5.4 & 4 (5) & 0 & - \\
HSC-z6PCC2 & Deep & ELAIS-N1 & 243.22 & 53.92 & 15.0 & 8.4 & 6 (8) & 0 & - \\ 
HSC-z6PCC8 & Deep & ELAIS-N1 & 243.89 & 54.42 & 9.7 & 5.4 & 4 (4) & 0 & - \\

HSC-z6PCC9 & Deep & DEEP2-3 & 351.30 & 0.03 & 9.7 & 5.4 & 4 (4) & 0 & - \\
HSC-z6PCC10 & Deep & DEEP2-3 & 351.95 & $-0.10$ & 9.7 & 5.4 & 4 (6) & 0 & - \\
HSC-z6PCC11 & Deep & DEEP2-3 & 352.72 & 0.60 & 9.7 & 5.4 & 4 (4) & 0 & - \\
HSC-z6PCC12 & Deep & DEEP2-3 & 352.84 & 0.91 & 9.7 & 5.4 & 4 (6) & 0 & - \\
HSC-z6PCC13 & Deep & DEEP2-3 & 352.97 & 0.08 & 9.7 & 5.4 & 4 (4) & 0 & - \\
HSC-z6PCC14 & Deep & DEEP2-3 & 353.45 & $-0.10$ & 9.7 & 5.4 & 4 (4) & 0 & - \\  \hline

	\multicolumn{10}{c}{$z=6.6$}\\ \hline
HSC-z7PCC9 & UD & SXDS & 34.62 & $-5.13$ & 6.6 & 4.6 & 4 (4) & 3 & 6.574 \\
HSC-z7PCC28 & UD & SXDS & 34.64 & $-4.56$ & 6.1 & 3.8 & 3 (3) & 5 & 6.537 \\

HSC-z7PCC11 & UD & COSMOS & 149.35 & 2.41 & 6.6 & 4.6 & 4 (4) & 0 & - \\
HSC-z7PCC15 & UD & COSMOS & 150.30 & 2.00 & 6.6 & 4.6 & 4 (6) & 0 & - \\
HSC-z7PCC16 & UD & COSMOS & 150.48 & 2.29 & 6.6 & 4.6 & 4 (8) & 0 & - \\

HSC-z7PCC1 & Deep & COSMOS & 148.96 & 1.02 & 10.5 & 7.2 & 6 (6) & 0 & - \\
HSC-z7PCC10 & Deep & COSMOS & 149.05 & 3.10 & 6.6 & 4.6 & 4 (4) & 0 & - \\
HSC-z7PCC2 & Deep & COSMOS & 149.40 & 1.03 & 8.5 & 5.9 & 5 (5) & 0 & - \\
HSC-z7PCC12 & Deep & COSMOS & 149.41 & 3.54 & 6.6 & 4.6 & 4 (4) & 0 & - \\
HSC-z7PCC13 & Deep & COSMOS & 149.67 & 2.79 & 6.6 & 4.6 & 4 (4) & 0 & - \\
HSC-z7PCC14 & Deep & COSMOS & 149.97 & 1.45 & 6.6 & 4.6 & 4 (6) & 0 & - \\
HSC-z7PCC3 & Deep & COSMOS & 150.95 & 2.78 & 8.5 & 5.9 & 5 (5) & 1 & 6.575 \\
HSC-z7PCC17 & Deep & COSMOS & 151.15 & 3.49 & 6.6 & 4.6 & 4 (4) & 0 & - \\
HSC-z7PCC18 & Deep & COSMOS & 151.16 & 3.13 & 6.6 & 4.6 & 4 (2) & 0 & - \\

HSC-z7PCC19 & Deep & ELAIS-N1 & 240.74 & 54.63 & 6.6 & 4.6 & 4 (4) & 0 & - \\
HSC-z7PCC4 & Deep & ELAIS-N1 & 241.27 & 54.4 & 8.6 & 5.9 & 5 (5) & 0 & - \\
HSC-z7PCC20 & Deep & ELAIS-N1 & 241.58 & 56.33 & 6.6 & 4.6 & 4 (4) & 0 & - \\
HSC-z7PCC21 & Deep & ELAIS-N1 & 241.92 & 55.66 & 6.6 & 4.6 & 4 (7) & 0 & - \\
HSC-z7PCC22 & Deep & ELAIS-N1 & 241.95 & 53.76 & 6.6 & 4.6 & 4 (4) & 0 & - \\
HSC-z7PCC5 & Deep & ELAIS-N1 & 242.31 & 56.4 & 8.6 & 5.9 & 5 (5) & 0 & - \\
HSC-z7PCC23 & Deep & ELAIS-N1 & 242.38 & 55.03 & 6.6 & 4.6 & 4 (4) & 0 & - \\
HSC-z7PCC24 & Deep & ELAIS-N1 & 242.47 & 53.48 & 6.6 & 4.6 & 4 (4) & 0 & - \\
HSC-z7PCC8 & Deep & ELAIS-N1 & 243.33 & 56.53 & 6.7 & 4.6 & 4 (4) & 0 & - \\
HSC-z7PCC25 & Deep & ELAIS-N1 & 243.52 & 56.03 & 6.6 & 4.6 & 4 (5) & 0 & - \\
HSC-z7PCC6 & Deep & ELAIS-N1 & 243.73 & 55.13 & 8.6 & 5.9 & 5 (5) & 0 & - \\
HSC-z7PCC7 & Deep & ELAIS-N1 & 243.93 & 54.35 & 8.6 & 5.9 & 5 (5) & 0 & - \\

HSC-z7PCC26& Deep & DEEP2-3 & 351.09 & $-0.77$ & 6.6 & 4.6 & 4 (4) & 0 & - \\
HSC-z7PCC27 & Deep & DEEP2-3 & 353.04 & 0.77 & 6.6 & 4.6 & 4 (4) & 0 & - 
\enddata
\tablecomments{
(1) object ID;
(2) layer;
(3) field; 
(4) right ascension of the center of the member LAEs (deg); 
(5) declination of the center of the member LAEs (deg); 
(6)-(7) highest $\delta$ and the significance level in the protocluster candidates;
(8) number of the HSC LAEs in a 0.07 deg radius from the center of the protocluster candidates;
(9) number of the spectroscopically-confirmed LAEs in 10 cMpc from the center of the protocluster candidates; 
(10) average redshift value of the spectroscopically-confirmed LAEs.
}
\label{tab: pcclist}
\end{deluxetable*}
\clearpage


\bibliographystyle{apj}
\bibliography{cite_higuchi2017}

\end{document}